\documentclass[final,5p,times,twocolumn]{elsarticle}
 \biboptions{comma,sort&compress}

\usepackage{graphicx}
\usepackage{here}
\usepackage{axodraw}
\def\beq{\begin{equation}}
\def\eeq{\end{equation}}
\def\bea{\begin{eqnarray}}
\def\eea{\end{eqnarray}}
\def\bq{\begin{quote}}
\def\eq{\end{quote}}

\def\nnb{\nonumber}
\def\ga{\left(}
\def\dr{\right)}

\def\lrar{\Longrightarrow}
																
\def\nnb{\nonumber}
\def\la{\langle}
\def\ra{\rangle}
\def\nin{\noindent}
\def\ba{\vspace*{-0.2cm}\begin{array}}
\def\ea{\end{array}\vspace*{-0.2cm}}

\def\b{$\bullet~$}
\def\als{\alpha_s}

\def\gg2{ \la\alpha_s G^2 \ra}
\def\gg3{g^3f_{abc}\la G^aG^bG^c \ra}
\def\ggg4{\la\als^2G^4\ra}

\def\calD{ {\cal D} }
\def\ftilde{\tilde f}

\def\therho{\theta\rho}

\def\frac#1#2{{#1\over#2}}

\def\calD{ {\cal D} }
\def\ftilde{\tilde f}

\def\therho{\theta\rho}

\journal{Physics Letters B}

\begin{document}

\begin{frontmatter}

%%
%%%%%%%%%%%%%%%%%%%%%%%%%%%%%%%%%%%%%%%%%%%%%%%%%
%\begin{document}

\title{ $\sigma$  and $f_0(980)$ substructures  from  $\gamma\gamma\to \pi\pi$, ~$J/\psi,\phi$
radiative and $D_s$ semi-leptonic decays}   
%\to\gamma G$ }

%\title{  Substructures of $\sigma(600)$  and $f_0(980)$   from  $\gamma\gamma\to \pi\pi$, ~$J/\psi,\phi\to \gamma f_0$
%and $D_s\to f_0 l\nu$}

 \author[label1]{G. Mennessier }
\ead{gerard.mennessier@lpta.univ-montp2.fr}

 \author[label1]{S. Narison\corref{cor1} }
   \address[label1]{Laboratoire
de Physique Th\'eorique et Astroparticules, CNRS-IN2P3,  
Case 070, Place Eug\`ene
Bataillon, 34095 - Montpellier Cedex 05, France.}
\cortext[cor1]{Corresponding author}
\ead{snarison@yahoo.fr}

 \author[label1,label3]{X.-G. Wang,\corref{cor2}}
  \address[label3]{Department of Physics, Peking University, Beijing 100871, China.}
\cortext[cor2]{China scholarship council fellow under contract n$^0$ 2009601139.}
\ead{wangxuangong@pku.edu.cn}

%\begin{document}

%\pagestyle{myheadings}
%\markright{ }
\begin{abstract}
\noindent
Using an improved ``analytic $K$-matrix model", we reconsider the extraction of the $\sigma\equiv f_0(600)$ and $f_0(980)$ $\gamma\gamma$ widths from  $\gamma\gamma\to \pi\pi $ scatterings data of Crystal Ball and Belle. 
Our main results are summarized in Tables \ref{tab:physical} and \ref{tab:width}.  The averaged $\sigma$  ``direct width" to $\gamma\gamma$ is 0.16(3) keV which confirms a previous result of \cite{MNO} and which  does neither favour a large four-quark / molecule nor a pure $\bar qq$ components.
%but an eventual large gluon component already expected from the analysis of the hadronic processes $\pi\pi\to \pipi/\bar KK$. 
The ``direct width" of the $f_0(980)$ of 0.28(2) keV is much larger than the four-quark  expectation but can be compatible with a $\bar ss$ or a gluonium component.  We also found that the rescattering part of each amplitude is relatively large indicating an important contribution of the meson loops in the determinations of the $\sigma$ and $f_0(980)$ $\gamma\gamma$ total widths. 
This is mainly due to the large couplings of the $\sigma$ and $f_0(980)$ to $\pi\pi$ and/or $\bar KK$, which can also be due to a light scalar gluonium with large OZI violating couplings but not necessary to a four-quark or molecule state.  Our average results for the total (direct+rescattering) $\gamma\gamma$ widths:  $\Gamma_\sigma^{tot}= 3.08(82)~{\rm keV},~  \Gamma_{f_0}^{tot}= 0.16(1)~{\rm keV}$  are comparable with the ones from dispersion relations and PDG values. Using the parameters from QCD spectral sum rules, we complete our analysis by showing that the production rates  of unmixed scalar gluonia $\sigma_B(1)$ and G (1.5-1.6) agree  with the data from 
%from  $\gamma\gamma\to \pi\pi$, ~$J/\psi,\phi\to \gamma f_0$ and $D_s\to f_0 l\nu$.
$J/\psi$, $\phi$ radiative and $D_s$ semi-leptonic decays.
%Some comments on a possible nature of the  $\sigma$ and $f_0(980)$ are given.
\end{abstract}


\end{frontmatter}

%\maketitle
%%%%%%%%%%%%
%\vspace*{-1.5cm}
\section{Introduction}
\vspace*{-0.25cm}
 \nin
%%%%%%%%%%%%
In previous series of papers \cite{MNO,KMN,MNW}, we have used an improved version of the K-matrix model originally proposed in \cite{MENES} for studying the hadronic and $\gamma\gamma$ couplings of the $\sigma/f_0(600)$ meson\,\footnote{Some other applications of the model have been discussed in \cite{PEAN,LAYSSAC}.} . We found that the ``direct" coupling of the $\sigma$ to $\gamma\gamma$ is more compatible with a large gluon component in its wave function rather than with a $\bar qq$ (too large $\gamma\gamma$ width) or four-quark (too small $\gamma\gamma$ width). More recently, we have extended the analysis for studying the hadronic couplings of the $\sigma/f_0(600)$ and $f_0(980)$ mesons
\cite{KMN,MNW}. We found an unexpected relatively large coupling of the $\sigma$ to $\bar KK$: $|g_{\sigma K^+K^-}|/|g_{\sigma\pi^+\pi^-}|=0.37(6)$\,\footnote{Analogous values have been obtained in \cite{MENES,OTHERS,ZHENG09} and in Fits 9 and 10 of \cite{ACHASOV06} though not favoured by \cite{ACHASOV06}.}, which disfavours its large $\pi-\pi$ molecule and four-quark components, while the large coupling of $f_0(980)$ to $\bar KK$: $|g_{fK^+K^-}|/|g_{f\pi^+\pi^-}|=2.59(1.34)$, excludes its pure $(\bar uu+\bar dd)$ content. These phenomenological observations go in lines with the fact that, in the $I=0$ channel, the gluon component is expected to play an essential r\^ole through the
scalar QCD anomaly (dilaton) \cite{VENEZIA,NSVZ,SNG0,SN06,LANIK,SCHEC,CHANO,OCHS,FRASCA,ARRIOLA} which manifests through the trace of the QCD energy momentum tensor:
%\,\footnote{This important property of QCD should be checked on the lattice.}:
\beq
\theta^\mu_\mu= \sum_{i=u,d,s} m_i(1+\gamma_m)\bar\psi_i\psi_i +{1\over 4}\beta(\alpha_s)G^2~,
\eeq
where $\gamma_m$ is the quark mass anomalous dimension; $\psi_i$ is the quark field; $\beta(\alpha_s)$ is the Gell-Mann-Low QCD $\beta$-function and $G^2$ is the gluon field strength.\\
 In this paper, we pursue the test of the nature of these scalar mesons by studying their couplings to $\gamma\gamma$ inside the energy region below 1.4 GeV, where new data on $\gamma\gamma\to\pi\pi$ from BELLE \cite{BELLE}
is available in addition to the old data of Crystal Ball \cite{CBALL} and MARK II \cite{MARK2}.
  %%%%%%%%%%%%%%%%%%%%%%%%
%\vspace*{-0.5cm}
%\vspace*{-0.45cm}
\section{The analytic K-matrix model for $\pi\pi\to\pi\pi/\bar KK
%\gamma\gamma
$}
\vspace*{-0.25cm}
\nin
%%%%%%%%%%%%%%%%%%%%%%%%
In so doing, we shall work with a specific analytic K-matrix model  originally introduced by the authors in Ref. \,\cite{MENES}, where one can separate the direct and rescattering $\gamma\gamma$ couplings,
which is not always feasible using dispersion relations. \\
In  this approach, the strong processes are described by a K-matrix model representing the amplitudes by a set of resonance poles and where
the dispersion relations in the multi-channel case 
can be solved explicitly. 
The  model can be reproduced by a set of Feynman diagrams, which are easily interpreted within the Effective Lagrangian approach, including 
resonance (bare) couplings to  $\pi\pi$ and $K\bar K$
 and (in the original model \cite{MENES}) 4-point $\pi\pi$ and $K\bar K$ interaction vertices
 which we have omitted for simplicity in \cite{MNO} and here.  
A subclass of bubble pion
 loop diagrams including resonance poles in the s-channel are resummed
 (unitarized Born).  In a previous work \cite{MNW}, we have discussed the approach for the case of : 
{\it 1~channel $\oplus$ 0~``bare" resonance (so-called $\lambda \Phi^4$ model)},  {\it 1~channel $\oplus$ 1~``bare" resonance (K-matrix pole)} and
{\it 2~channels $\oplus$ 2~``bare" resonances} 
and we have restricted to the $SU(3)$ symmetric shape function. We have introduced a real analytic form factor {\it shape function}, which takes
explicitly into account left-handed cut singularities for the strong interaction amplitude,
and which allows a more flexible parametrisation of the $\pi \pi\to\pi\pi/\bar KK$ data. 
In our low energy approach, the shape function can be conveniently
approximated by\,\footnote{Here and in the following $\sigma_D$ is negative and is opposite in sign with the
one used in our previous works \cite{MNO,KMN,MNW}.}:
  \beq
f_P(s)=\frac{s-s_{AP}}{s-\sigma_{DP}} \label{formfactor}~,~~P\equiv \pi,~K~,
\eeq    
which multiplies the scalar meson couplings to $\pi\pi/\bar KK$. In this form, the {\it shape function} allows for an Adler  zero at $s=s_{AP}$
and a pole for $\sigma_{DP}<0$ simulating the left hand cut. 
%%%%%%%%%%%%%%%%%%%%%%%%%%%
%\vspace*{-0.3cm}
\subsection*{\b \bf 1~channel $\oplus$ 1~``bare" resonance}
%\vspace*{-0.25cm}
%%%%%%%%%%%%%%%%%%%%%%%%%%%
\nin
Let's first illustrate the method in this simple case. 
The unitary $PP$ amplitude is then written as:
\beq
  T_{PP}(s) = \frac{G_P f_P(s)}{s_R-s -  G_P \ftilde_P (s)} = \frac{G_P f_P(s)}{\calD_P(s)}~,
\label{tpipi}
\eeq 
where $T_{PP}=e^{i\delta_P}\sin\delta_P/\rho_P(s)$ with 
 $ \rho_P(s) =({1 - 4 m^2_P/s})^{1/2}$;  $G_P=g_{\sigma P,B}^2$
 are the bare coupling squared and :
\beq
{\rm {Im}}~ \calD_P = {\rm{Im}} ~ (- G_P \ftilde_P) = - (\therho_P) G_P \ f_P~,  
\label{eq:imaginary}
\eeq
with: $(\theta\rho_P)(s)=0$ below and $(\theta\rho_P)(s)=\rho_P(s)$
above threshold $s=4m_P^2$. The ``physical" couplings are defined from the residues,
with the normalization:
\beq
g_{\sigma P}^2\equiv g^2_{\sigma PP}/(16\pi)~:~~~ ~\Gamma_{\sigma\to\pi\pi}={|g_{\sigma PP}(M^2_\sigma)|^2\over 16\pi M_\sigma}\rho_P(M_\sigma^2).
\eeq
The amplitude near the pole $s_0$ where $ {\cal D}_P(s_0)=0$ and
${\cal D}_P(s)\approx {\cal D'}_P(s_0) (s-s_0)$ is:
\beq
  T_{PP}(s)\sim \frac{g_{\sigma P}^2}{s_0-s}; \qquad g_{\sigma P}^2=\frac{G_P
f_P(s_0)}{-\calD'_P(s_0)}~.
\label{eq:gpi2}  
\eeq
The real part of $\calD_P$ is obtained from a dispersion relation with
subtraction at $s=0$ and one obtains:
\beq
\ftilde_P(s) = \frac{2}{\pi} \Big{[} h_P(s) \ -h_P(0) \Big{]}~,
\label{eq:ftilde}
\eeq
with: $h_P(s) =f_P(s) \tilde L_{s1}(s)$--$(\sigma_{NP}/(s-\sigma_{DP}))\tilde L_{s1}(\sigma_{DP})$, 
 $\sigma_{NP}$ is the residue of $f_P(s)$ at $\sigma_{DP}$ and: $\tilde L_{s1}(s) =  
 \big{[}\ga{s - 4 m_P^2}\dr/{m_P^2} \big{]}
\tilde L_1(s,m_P^2)$ where: $\tilde L_1$ from \cite{MENES}. 
%%%%%%%%%%%%%%%%%%%%%%%%%%%%
%\vspace*{-0.3cm}
\subsection*{\b \bf Generalization to 2 channels $\oplus$ 2 ``bare" resonances}
%%%%%%%%%%%%%%%%%%%%%%%%%%%%
 \nin
In \cite{KMN,MNW}, we have generalized the previous case to the one of 2 channels $\oplus$ 2 ``bare" resonances. 
%We show in Table \ref{tab:param} the  results obtained in \cite{MNW} from 3 different sets of the experimental inputs which we expect to span the whole space of parameters. 
  %%%%%%%%%%%%%%%%%%%%%%%%
  %\vspace*{-0.5cm}
\section{The $\gamma\gamma\to\pi\pi$ process}
%\gamma\gamma
\vspace*{-0.25cm}
\nin
%%%%%%%%%%%%%%%%%%%%%%%%
%%%%%%%%%%%%%%%%%%%%%%%%%%%%
%\vspace*{-0.3cm}
\subsection*{\b \bf Expression and normalization of the amplitudes  %for \boldmath$\gamma\gamma\to \pi\pi$ 
}
%%%%%%%%%%%%%%%%%%%%%%%%%%%%
 \nin
 The amplitude
$\gamma(q_1,\epsilon_1)+\gamma_2(q_2,\epsilon_2)\rightarrow
\pi(p_1)+\bar{\pi}(p_2)$ of mass $m_\pi$ can be written in terms of
the invariants \footnote{We use the same normalization as \cite{MENES}.}: \begin{eqnarray}\label{I1I2}
A&=&I_1 A_1+I_2 A_2\ ,\nonumber\\
I_1&=&(\epsilon_1\cdot\epsilon_2)-(\epsilon_1\cdot
q_2)(\epsilon_2\cdot q_1)/(q_1\cdot q_2)\ ,\nonumber\\
I_2&=&(\epsilon_1\cdot\Delta)(\epsilon_2\cdot\Delta)(q_1\cdot
q_2)-(\epsilon_1\cdot q_2)(\epsilon_2\cdot\Delta)(q_1\cdot\Delta)\nnb\\
&&-(\epsilon_2\cdot q_1)(\epsilon_1\cdot\Delta)(q_2\cdot\Delta)\nonumber\\
&&+(\epsilon_1\cdot q_2)(\epsilon_2\cdot
q_1)(q_1\cdot\Delta)(q_2\cdot\Delta)/(q_1\cdot q_2)\ ,
\end{eqnarray}
with  $\Delta=p_1-p_2$. Helicity $\lambda=0$ and $\lambda=2$ amplitudes are denoted by F and
G, which, in terms of partial wave amplitudes, read respectively: 
\begin{eqnarray}\label{FG}
F&=&A_1-s(s/4-m_\pi^2)\sin^2\theta A_2\nnb\\
&=&\sum_{even
J\geq0}(2J+1)f^{J0}(s)d^{J}_{00}(\theta)\, , \nonumber\\
G&=&s(s/4-m_\pi^2)\sin^2\theta A_2\nnb\\
&=&\sum_{even J\geq2}(2J+1)g^{J2}(s)d^{J}_{20}(\theta)4/\sqrt{6}\ .
\end{eqnarray}
$d_{20}^J$ and $d_{00}^J$ are the usual $d$-functions normalized as in PDG \cite{PDG}, while $\theta$ is the scattering angle between $\overrightarrow{p}$ and
$\overrightarrow{q}$, which can be expressed  in
terms of s and t as:
\begin{equation}\label{costheta}
\cos\theta=\frac{2t+s-2m_{\pi}^2}{\sqrt{s(s-4m_{\pi}^2)}}\ .
\end{equation}
For unpolarized photons, the cross section reads: 
\begin{equation}
\frac{d\sigma}{d\Omega}=\frac{2}{s}\sqrt{1-\frac{4m_\pi^2}{s}}(|F|^2+|G|^2)\
,
\end{equation}
where the $\cos\theta$ integration should be done 
from 0 to 1 for the neutral and from -1 to 1 for the charged cases.
In the following analysis, we find convenient to express the charged $F_C$ and neutral $F_N$ amplitudes in terms of the $I=0$ and $I=2$ isospin ones \footnote{We use the same convention as \cite{MENES} where $F_C$ is opposite in sign with \cite{OLLER08}.}:
\begin{eqnarray}\label{FCN}
F_C&=&\sqrt{2\over 3}\ga F^{I=0}+ {1\over \sqrt{2}}F^{I=2}\dr\
,\nonumber\\
F_N&=&-\sqrt{2\over 3}\ga F^{I=0}-\sqrt{2}F^{I=2}\dr\ .
\end{eqnarray}
corresponding to the following $|\pi\pi\ra$ states:
\begin{eqnarray}
|\pi\pi,I=0>&=&\sqrt{2\over 3}\ga |\pi^+\pi^-\ra-\frac{1}{2}|\pi^0\pi^0\ra\dr\
, \nonumber\\
|\pi\pi,I=2>&=&\sqrt{1\over 3}\ga |\pi^+\pi^-\ra+|\pi^0\pi^0\ra\dr\
.
\end{eqnarray}
%%%%%%%%%%%%%%%%%%%%%%%%%%%%
%%%%%%%%%%%%%%%%%%%%%%%%%%%%%%%%%%%%%%%%%%%
%\vspace*{-0.3cm}
\subsection*{\b \bf The example of one pion exchange for \boldmath$\gamma\gamma\to \pi^+\pi^-$ 
}
%%%%%%%%%%%%%%%%%%%%%%%%%%%%
 \nin
 %%%%%%%%%%%%%%%%%%%%%%%%%%%%%%%%%%
%\subsection{One Pion Exchange}
The expression of the Born term amplitude due to one pion
exchange  reads \footnote{Here and in the following, we use the same normalization as in \cite{MENES} and we use the gauge conditions: $\epsilon_iq_i=\epsilon_iq_j=0~: i,j=1,2$.}:
\begin{equation}\label{TBpi}
T^{B}_\pi=\frac{\alpha}{2}\Big{[}\epsilon_1\cdot\epsilon_2-\frac{2(\epsilon_1\cdot
p_1)(\epsilon_2\cdot p_1)}{t-m_{\pi}^2}-\frac{2(\epsilon_1\cdot
p_1)(\epsilon_2\cdot p_1)}{u-m_{\pi}^2}\Big{]}
\end{equation}
from which one can deduce the helicity amplitudes: 
\begin{eqnarray}\label{FGpi}
F^{B}_\pi&=&\frac{\alpha}{2}\frac{m_{\pi}^2s}{(t-m_{\pi}^2)(u-m_{\pi}^2)}\ ,\nonumber\\
G^{B}_\pi&=&\frac{\alpha}{2}\frac{t^2+ts-2m_{\pi}^2
t+m_{\pi}^4}{(t-m_{\pi}^2)(u-m_{\pi}^2)}\ .
\end{eqnarray}
%%%%%%%%%%%%%%%%%%%%%%%%%%%%%%%%%%%%%%%%%%%
%\vspace*{-0.5cm}
%\section{$\gamma\gamma\to\pi\pi$ for 1 channel $\oplus$ 1 resonance below 0.7 GeV}
%\gamma\gamma
%\vspace*{-0.25cm}
%\nin
%%%%%%%%%%%%%%%%%%%%%%%%%%%%%%%%%%%%%%%%%%%
%%%%%%%%%%%%%%%%%%%%%%%%%%%%
%\vspace*{-0.3cm}
\subsection*{\b \bf Amplitudes  for 1 channel $\oplus$ 1 resonance below 0.7 GeV %for \boldmath$\gamma\gamma\to \pi\pi$ 
}
%%%%%%%%%%%%%%%%%%%%%%%%%%%%
%\end{document}
 \nin
 {\it -- The isospin I=0 channel}:
starting from the
 S wave amplitude in Eq.~(\ref{tpipi}),  we derive the amplitude $T_\gamma^{(I)}$ for 
the electromagnetic process for isospin $I=0$ as:
\beq
 T_{\rm \gamma}^{(0)} = 
  \sqrt{\frac{2}{3}} \alpha \left(f_P^B + G_P\ \frac{\ftilde_P^B}{\calD_P}\right) + 
  \alpha \frac{\cal P}{\calD_P}~. \label{elmamp} 
\eeq
Here the contribution from the Born term of $\gamma\gamma\to \pi^+\pi^-$
is given by $f_P^B=2L_1$ as defined
in \cite{MENES}, a real analytic function in the $s$ plane 
with left cut $s\le 0$. The function 
$\ftilde_P^B$ represents $\pi\pi$ rescattering; 
it is regular for $s < 4 m_\pi^2$ but has a right cut 
for $s \ge 4 m_\pi^2$ with:
\beq
{\rm{Im}}~\ftilde_P^B (s+i\epsilon) =  (\therho \ f_P \ f_P^B)(s)~,
\eeq
which vanishes at $s=0$.
With this definition the Watson theorem is fulfilled, i.e. the phase of 
$ T^{(0)}_\gamma$ is the same as the one of the elastic amplitude ${\cal D}_P^{-1}$
in Eq. (\ref{tpipi}). The real part is derived from a dispersion 
relation with subtraction at $s=0$ for $\ftilde_P^B$ to satisfy the Thomson limit,
and has a representation similar to the one in Eq. (\ref{eq:ftilde}), 
but by replacing
$(\tilde f_P,h_P)$ by $(\tilde f^B_P,h_P^B)$.
The function $h_P^B$ is defined as $h_P$ below Eq. (\ref{eq:ftilde}) but with 
$\tilde L_1$ replaced by $-\tilde L_1^2$ everywhere. It
vanishes at $s=\sigma_{DP}$ and $ \ftilde_P^B(s)$
is regular at this point. 
Finally, the polynomial $\cal P$ reflects the ambiguity from the dispersion
relations and is set here to ${\cal P} = s F_\gamma\sqrt{2}$. It represents the direct
coupling of the resonance to $\gamma\gamma$. The residues 
 at the pole $s_0$ of the
rescattering  and direct contributions to $ T^{(0)}_\gamma$ in Eq.~(\ref{elmamp}), 
respectively, are obtained as: 
 \beq
g^{\rm resc}_{\sigma\gamma}  g_{\sigma\pi} =  \sqrt{\frac{2}{3}} \alpha 
 \frac{ G\ftilde_P^B(s_0)}{-\calD_P'(s_0)}; ~~~
    g^{\rm dir}_{\sigma\gamma} g_{\sigma\pi} = \alpha\frac{ s_0 F_\gamma \sqrt{2}}{-\calD_P'(s_0)},
 \label{eq:fgamma}
 \eeq
from which one can deduce the branching ratio:
\beq
{\Gamma_{\sigma\to\gamma\gamma}\over \Gamma_{\sigma\to\pi\pi}}
\simeq {1\over \vert \rho(s_0)\vert }\Big{\vert} {g_{\sigma\gamma} \over g_{\sigma\pi}
}\Big{\vert}^2
\simeq {2\alpha^2\over \vert \rho(s_0)\vert }\Big{\vert} {s_0\over Gf_P(s_0)}\Big{\vert} ^2 F_\gamma^2~.
\label{eq:gamwidth}
\eeq
{\it -- The isospin I=2 channel}:
similarly, we parametrize
the  $I=2$ $S$-wave amplitude $T^{(2)}_0$ 
by introducing the shape function $f_2$:
\beq
 T^{(2)}_0= {\Lambda f_2(s) \over 1-\Lambda \tilde f_2(s)},
~~f_2(s)={s-s_{A2}\over (s - \sigma_{D1})(s - \sigma_{D2})}~,
\label{eq:t2f2}
\eeq
and obtain:
\beq 
T^{(2)}_\gamma={\alpha\over \sqrt{3}}\ga f_2^B+ {\Lambda \tilde f^B_2(s) \over 1-\Lambda \tilde f_2(s)}\dr~,
\label{eq:i2amplitude}
\eeq 
where $f_2^B=f^B_P$ and:
$
{\rm Im} \tilde{f_2}(s)  = (\theta  \rho) f_2(s),
~{\rm Im} \tilde{f^B_2}(s) = (\theta \rho~ f_2~ f_2^B)(s).
$
These amplitudes are again both subtracted at $s=0$ as in case of $I=0$ 
and one finds in analogy:
\beq
\tilde f_2(s) = \frac{2}{\pi}  \Big{[} h_2(s)-h_2(0)  \Big{]}~,
\label{f2tilfct}
\eeq
where:
$h_2(s) =
f_2(s)\tilde L_{s1}(s)-
({\sigma_{N1}}/({s-\sigma_{D1}}))
~\tilde L_{s1}(\sigma_{D1})
 $ $  - ( {\sigma_{N2}}/({s-\sigma_{D2}}))\tilde L_{s1}(\sigma_{D2})$;
$\sigma_{N1},\sigma_{N2}$ are the residues  of $f_2(s)$ at 
$\sigma_{D1},\sigma_{D2}$ and $\tilde f_2^B(s)$ is 
defined as $\tilde f_2(s)$ in Eq. (\ref{f2tilfct})
but with $\tilde L_{1}$ replaced by  $-\tilde L_{1}^2$.  
The cross sections for the $\pi\pi$ and $\gamma\gamma$ 
scattering processes are obtained 
from the previous expressions of the amplitudes.
%as described in the appendices of \cite{MENES}, 
%where the $\cos\theta$ integration should be done 
%from 0 to 1 for the neutral and from -1 to 1 for the charged cases.
%\footnote{The range of integration may have been misused in \cite{HARJES}.}.
%%%%%%%%%%%%%%%%%%%%%%%%%%%%
%\vspace*{-0.3cm}
\subsection*{\b \bf Results of the analysis}
\nin 
%%%%%%%%%%%%%%%%%%%%%%%%%%%%
In doing the analysis for the case of elastic $\pi\pi\to\pi\pi$ scattering and $\gamma\gamma\to \pi\pi$
below 700 MeV where the Crystal Ball \cite{CBALL} and MARK II \cite{MARK2} data have been used, we have obtained in \cite{MNO} the results summarized in Table \ref{tab:width}. 
  %%%%%%%%%%%%%%%%%%%%%%%%%%%%%%%%%%%%%%
%\vspace*{-0.5cm}
\section{Extension of the $\gamma\gamma\to\pi\pi$ analysis below 1.09 GeV }
\vspace*{-0.25cm}
\nin
%%%%%%%%%%%%%%%%%%%%%%%%%%%%%%%%%%%%%%%
In this paper, we extend the previous analysis by including $\pi\pi$ and $\bar KK$ loops and work in the region just above the $\bar KK$ threshold (the minimal $\chi^2/ndf$ of our fit is obtained for $\sqrt{s}=1.09$ GeV), where the $\sigma(600),~f_0(980)$ contributions are dominant. 
%%%%%%%%%%%%%%%%%%%%%%%%%%%%
%\vspace*{-0.3cm}
\subsection*{\b \bf S-waves masses and hadronic couplings}
\nin 
%%%%%%%%%%%%%%%%%%%%%%%%%%%%
In this case with 2 resonances $\oplus$ 2 channels, the hadronic masses and couplings of the $\sigma(600)$ and $f_0(980)$ and the corresponding values of the ``bare parameters" of the model are given in Tables 2 and 3 of  \cite{MNW} which we shall use in the extraction of their $\gamma\gamma$ couplings.   The values  of the masses are (in units of MeV):
\beq
M_\sigma=452(12)-{\rm i} 260(15),~M_f=981(34)-{\rm i} 18(11),
\eeq
 and the ratios of the hadronic couplings are:
 \beq
{ |g_{\sigma K^+K^-}|\over |g_{\sigma\pi^+\pi^-}|}=0.37(6)~,~~~~{|g_{fK^+K^-}|\over |g_{f\pi^+\pi^-}|}=2.59(1.34)~.
 \eeq
  %%%%%%%%%%%%%%%%%%%%%%%%%%%%%%%%%%%%%%
%\vspace*{-0.5cm}
%\section{Analysis of $\gamma\gamma\to\pi\pi$ below 1.4 GeV}
%\vspace*{-0.25cm}
%\nin
%%%%%%%%%%%%%%%%%%%%%%%%%%%%%%%%%%%%%%%
%%%%%%%%%%%%%%%%%%%%%%%%%%%%
%\vspace*{-0.3cm}
%\subsection*{\b \bf Expression of the amplitudes and cross-section}
%\nin 

%%%%%%%%%%%%%%%%%%%%%%%%%%%%
%\vspace*{-0.5cm}
\subsection*{\b \bf D-wave mass and hadronic couplings}
\nin 
%%%%%%%%%%%%%%%%%%%%%%%%%%%%
\begin{figure}[h]%
\begin{center}%
%\vspace{2cm}
\includegraphics[width=6cm]{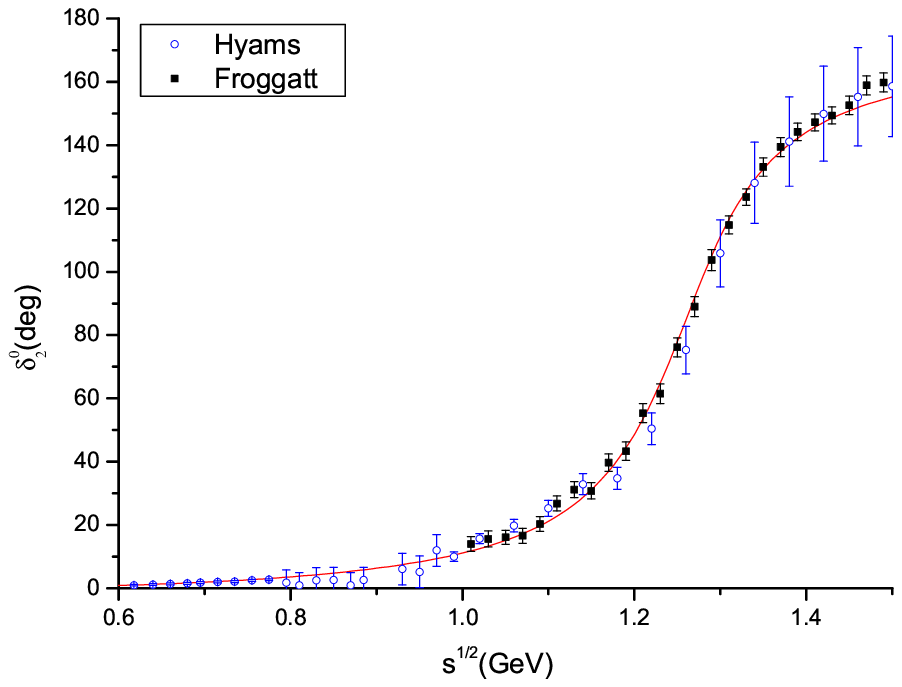}
\includegraphics[width=6cm]{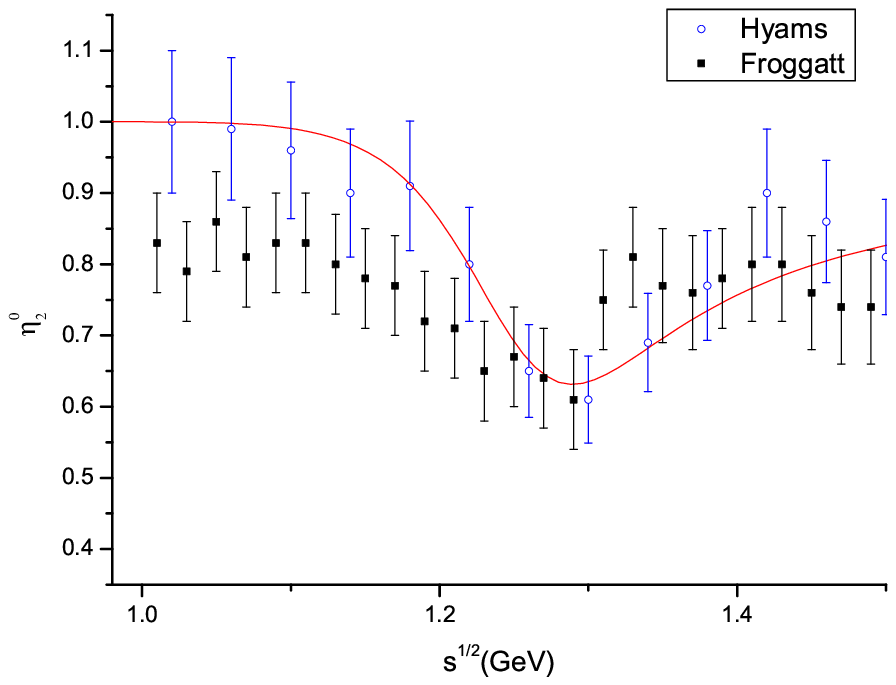}
\vspace*{-0.3cm}
%\vspace{-2cm}
\caption{\footnotesize  a) $I=0$ D wave $\pi\pi$ phase; b)
inelasticity. Experimental data are taken from~\cite{FP,HYAMS73}.}
\label{Dwave-fit}
\end{center}%
\end{figure}
\nin
%%%%%%%%%%%%%%%%%%%%%%%%%%%%
In the energy region where we shall work below 1.09 GeV, the $D$-wave contribution can be also 
important. However, an accurate parametrization of the $D$-wave contribution is not available. Assuming that it is dominated by the $f_2(1270)$, we extract its complex pole position and hadronic couplings and the corresponding ``bare parameters" of the model from the fit of the $I=0,~J=2$ phase shift measured in \cite{FP} and \cite{HYAMS73}. In so doing, we parametrize, as in \cite{MENES}, the $I=0$ D-wave $\pi\pi\to\pi\pi/K\bar{K}$ scattering
amplitudes:
\begin{equation}
T^{J=2}_{\pi K}=g_\pi g_K
\ga\frac{s-4m_\pi^2}{s-s_0}\dr\ga\frac{s-4m_K^2}{s-s_0}\dr
%(\frac{s-4m_j^2}{s-s_0})
\mathcal
{D}_{J=2}^{-1}\ ,
\end{equation}
where $s_0$ is a pole on the negative real axis to simulate left
hand singularity, and:
\begin{equation}
\mathcal {D}_{J=2}(s)=s_R-s-\sum_{P=\pi,K}g_{P}^2 C(s,m_{P})\ .
\end{equation}
The function $C(s,m_P)$ satisfies:
\begin{equation}
\mathrm{Im}C(s,m_P)=\rho_P(s)\ga\frac{s-4m_P^2}{s-s_0}\dr^2\ ,
\end{equation}
with $\rho_P(s)=\sqrt{1-4m_P^2/s}$ the phase-space function. The
real part of $C(s,m_P)$ can be obtained from dispersion relation:
\bea
C(s,m_P)&=&\frac{2}{\pi}\Bigg{[}\ga \frac{s-4m_P^2}{s-s_0}\dr^2\ga\frac{s-4m_P^2}{m_P^2}\tilde{L}_1(s)+1\dr\nnb\\
&&-\frac{s(2s_0-s)}{(s-s_0)^2}A-\frac{s}{s-s_0}B\Bigg{]}\, ,
\eea
where the last poles are adjusted to cancel the poles at $s=s_0$ and \footnote{Note the extra factor 1/2 in $B$ compared to the one in \cite{MENES}.}:
\begin{eqnarray}
A&=&\ga\frac{s_0-4m_P^2}{s_0}\dr^2\Big{[}\frac{s_0-4m_P^2}{m_P^2}\tilde{L}_1(s_0)+1\Big{]}~,\nonumber\\
B&=&\ga\frac{s_0-4m_P^2}{s_0}\dr^2\Big{[}2\frac{s_0+m_P^2}{m_P^2}\tilde{L}_1(s_0)
%\nnb\\&&
+\frac{s_0+4m_P^2}{s_0-4m_P^2}+\frac{1}{2}\Big{]}.\nnb\\
\end{eqnarray}
%\end{document}
The resulting values of the bare parameters are  given in Table \ref{tab:D-wave}, from which we derive the pole position in the 2nd sheet and the residues  of $f_2(1270)$ \footnote{There is also pole $(1.244-{\rm i}0.095)$ GeV in the 3rd sheet.}. 
Compared with the PDG data\,\footnote{Notice that the data of the inelasticity are not quite good which induces a relatively bad $\chi^2/{ndf}=132.6/95=1.4$.}:
%and which we have not tried to improve because the 
% in Fig \ref{Dwave-fit} as the 
% $D$-wave contribution remains a correction in the extraction of the $S$-wave parameters which is the main motivation of this paper.}:
\beq
M_{f_2}=1.27~ {\rm GeV~~~ and~~~}\Gamma_{f_2\to\pi\pi}=156.9^{+3.8}_{-1.2}~{\rm  MeV}~, 
\eeq
one can notice that the pole position and the $\pi\pi$ width are well reproduced. 
%This fact should be taken into account when  extracting the $\gamma\gamma$ width of the scalar mesons. 
%%%%%%%%%%%%%%%%%
\vspace*{-0.5cm}
{\footnotesize
\begin{table}[hbt]
\setlength{\tabcolsep}{0.2pc}
 \caption{\scriptsize    Values in GeV$^{d}~(d=1,2)$ of the bare and physical parameters of the $D$-wave }
 \footnotesize
\begin{tabular}{lllll}
%&\\
\hline
%\hline
%\\
&\\
$s_r=2.06$&$s_0=-1.25$&$g_\pi=0.96$ &$g_K=1.38$  \\
%\\
%\hline
%2.064&-1.248&0.961&1.377\\
&\\
%\hline
%\\
%&\\
$M_{f_2}=1.27-{\rm i} 0.07$&$g_{f_2\pi\pi}=0.48-{\rm i} 0.03$&$g_{f_2\bar KK}=0.28-{\rm i} 0.06$&\\
%\hline
%1.27-{\rm i} 0.07&0.48-{\rm i} 0.03&0.28-{\rm i} 0.06&\\
&\\
\hline
\end{tabular}
\label{tab:D-wave}
\end{table}
}
%%%%%%%%%%%%%%%%%%%%%%%%%%%%
%\vspace*{-0.3cm}
\subsection*{\b \bf The vector meson contributions}
\nin 
%%%%%%%%%%%%%%%%%%%%%%%%%%%%
In the energy-region where we shall work, exchange of vector
mesons $V\equiv\rho,~\omega,~K^{*+},~K^{*0}$ in the t-channel can become important. We introduce their couplings
to $\gamma\gamma$ via the effective interaction:
\beq
{\cal L}_{V\pi\gamma}={e\over 4}\epsilon^{\mu\nu\rho\sigma}\sum_V h_V~\pi~ V_{\mu\nu}F_{\rho\sigma}~,
\eeq
where $V_{\mu\nu}=\partial_\mu V_\nu-\partial_\nu V_\mu$ (resp. $F_{\mu\nu}=\partial_\mu F_\nu-\partial_\nu F_\mu$) are the vector (resp. electromagnetic) field strengths; $\pi$ is the pion field. The coupling $h_V$ is normalized as: 
\begin{equation}
\Gamma(V\rightarrow \pi\gamma)=\frac{\alpha}{24} h_V^2M_V^3\ga1-{m_\pi^2\over M_V^2}\dr^3 \ .
\label{eq:radiative}
\end{equation}
Using the standard vector form of the vector propagator,
the Born contribution to the $\gamma\gamma\to\pi^+\pi^-$ amplitude due to the vector meson exchange is:
%\scriptsize
\begin{eqnarray}
T_{V}^B&&=\frac{\alpha h_V^2}{16}\Big{[}\nnb\\
&&
\frac{(\epsilon_1\cdot\epsilon_2)[t(s-u)+m_{\pi}^4]-2s(\epsilon_1\cdot
p_1)(\epsilon_2\cdot
p_1)}{t-M_V^2}+\nonumber\\
&&
\frac{(\epsilon_1\cdot\epsilon_2)[u(s-t)+m_{\pi}^4]-2s(\epsilon_1\cdot
p_1)(\epsilon_2\cdot p_1)}{u-M_V^2}\Big{]},
\end{eqnarray}
from which we deduce the helicity amplitudes: 
\begin{eqnarray}
 F^{B}_V&=&\frac{\alpha h_V^2}{16}\Big{[}\frac{ts}{t-M_V^2}+\frac{us}{u-M_V^2}\Big{]}\ ,\nonumber\\
  G^{B}_V&=&\frac{\alpha
h_{V}^2}{16}\Big{[}\frac{t^2+ts-2m_{\pi}^2
t+m_{\pi}^4}{t-M_{V}^2}
\nnb\\&& 
+\frac{u^2+us-2m_{\pi}^2
u+m_{\pi}^4}{u-M_{V}^2}\Big{]}\ .
\end{eqnarray}
The results agree with the ones in the different literature (see e.g. \cite{MENES,GASSER94,OLLER08}). 
%%%%%%%%%%%%%%%%%%%%%%%%%%%%
%\vspace*{-0.3cm}
\subsection*{\b \bf The \boldmath$B(1^{+-})$ axial-vector contributions}
\nin 
%%%%%%%%%%%%%%%%%%%%%%%%%%%%
Their contributions can be introduced via the lowest order effective coupling \cite{GASSER94}:
\begin{equation}
\mathcal
{L}_{B\pi\gamma}=3eh_BF_{\mu\nu}Tr(\bar{B}^{\mu}\{Q,\partial^{\nu}\pi\})\
,
\end{equation}
with:
\begin{equation}
\bar{B}_{\mu}=\left(
             \begin{array}{cc}
               \frac{b_1^0}{\sqrt{2}}+\frac{h_1(1170)}{\sqrt{2}} & b_1^+(1235) \\
               b_1^- (1235)& -\frac{b_1^0}{\sqrt{2}}+\frac{h_1(1170)}{\sqrt{2}} \\              
               % K_{1B}^- & \bar{K}_{1B}^0 & h_1(1380) \\
             \end{array}
           \right)_{\mu}\ ,
\end{equation}
and $h_B$ is normalized as in Eq. (\ref{eq:radiative}). %\begin{equation}
%\Gamma(B\rightarrow \pi\gamma)=\frac{\alpha h_B^2}{3}\ga1-{m_\pi^2\over M_B^2}\dr^3\ .
%\end{equation}
One can deduce the amplitude:
\begin{eqnarray}
T_{B}^B&=&\frac{\alpha h_B^2}{16}\Big{[}\frac{(\epsilon_1\cdot\epsilon_2)(t-m_{\pi}^2)^2-2s(\epsilon_1\cdot
p_1)(\epsilon_2\cdot
p_1)}{t-M_B^2}\nonumber\\
&&
+\frac{(\epsilon_1\cdot\epsilon_2)(u-m_{\pi}^2)^2-2s(\epsilon_1\cdot
p_1)(\epsilon_2\cdot p_1)}{u-M_B^2}\Big{]}\ ,
\label{eq:amplitude_axial}
\end{eqnarray}
%The constant D is determined from the
%$b_1(1230)\rightarrow\gamma\pi^+$ decay
%\begin{equation}
%D^2=\frac{M_{b_1}^3
%\Gamma(b_1\rightarrow\gamma\pi^+)}{(M_{b_1}^2-m_{\pi}^2)^3}\frac{216f_{\pi}^2}{\alpha}\,
%\end{equation}
Then, the helicity amplitudes are:
\begin{eqnarray}
%A_1(b_1)&=&\frac{\alpha
%h_{b_1}^2}{8}\frac{1}{2}(\frac{(t-m_{\pi}^2)^2}{t-M_{b_1}^2}+\frac{(u-m_{\pi}^2)^2}{u-M_{b_1}^2})\
%,\nonumber\\
%A_2(b_1)&=&-\frac{\alpha
%h_{b_1}^2}{8}\frac{1}{2}(\frac{1}{t-M_{b_1}^2}+\frac{1}{u-M_{b_1}^2})\
%,\nonumber\\
F^{B}_{b_1}&=&-\frac{\alpha
h_{b_1}^2}{16}\Big{[}\frac{ts}{t-M_{b_1}^2}+\frac{us}{u-M_{b_1}^2}\Big{]}\ ,\nonumber\\
G^{B}_{b_1}&=&\frac{\alpha
h_{b_1}^2}{16}\Big{[}\frac{t^2+ts-2m_{\pi}^2
t+m_{\pi}^4}{t-M_{b_1}^2}\nnb\\
&&+\frac{u^2+us-2m_{\pi}^2
u+m_{\pi}^4}{u-M_{b_1}^2}\Big{]}\ .
\end{eqnarray}
%%%%%%%%%%%%%%%%%%%%%%%%%%%%
%\vspace*{-0.3cm}
\subsection*{\b \bf The \boldmath$a_1(1^{++})$ axial-vector contribution}
\nin 
%%%%%%%%%%%%%%%%%%%%%%%%%%%%
We describe the $a_1(1^{++})$ in the same way as the $b_1(1^{+-})$ meson\,\footnote{A tensor formulation of the axial-vector meson has been proposed in \cite{DERAFAEL} where the form
of the propagator differs from the standard one. If we use this  propagator, we reproduce the expression of the amplitude given in \cite{OLLER08} where an extra contact term $s(\epsilon_1.\epsilon_2)$ 
is added in Eq. (\ref{eq:amplitude_axial}). We shall see in our analysis that the presence of this term would decrease the strength of the direct coupling of the scalar resonance but the total = direct+rescattering
contribution remains almost unchanged. This term might be absorbed by some  other counter terms of the complete effective lagrangian. 
%We thank J. Gasser, E. de Rafael and J. Oller for some communications and discussions about this point. 
}, where
$h_{b_1}$ is simply replaced by $h_{a_1}$, which can either be determined from
the $a_1\to\pi\gamma$ width or from the ChPT coupling constants\,\cite{DERAFAEL,OLLER08}:
\begin{equation}
h_{a_1}^2
%=\frac{F_A^2}{f_{\pi}^2M_A^2}
=\frac{4(L_9+L_{10})}{f_{\pi}^2}=0.656~\mathrm{GeV}^{-2}\ ,
\end{equation}
where $f_{\pi}=92.4~\mathrm{MeV}$ is the pion decay constant. 
%%%%%%%%%%%%%%%%%%%%%%%%%%%%
%\vspace*{-0.3cm}
\subsection*{\b \bf The size of the different radiative couplings}
\nin 
%%%%%%%%%%%%%%%%%%%%%%%%%%%%
These can extracted from the data and using $SU(3)$ symmetry relations and read in units of GeV$^{-1}$:
\begin{eqnarray}
h_{\rho}&=&0.82,~ ~h_{\omega}=2.39, ~~
%\nonumber\\
h_{K^{*+}}=0.83,~~ h_{K^{*0}}=1.27,\nonumber\\
h_{a_1}&=&0.81,\ \  ~~h_{b_1}=0.65,\ \  ~~h_{h_1}=3h_{b_1}\ .
\label{eq:radcoupling}
\end{eqnarray}

  %%%%%%%%%%%%%%%%%%%%%%%%%%%%%%%%%%%%%%
%\vspace*{-0.5cm}
\section{K-matrix model analysis of $\gamma\gamma\to\pi\pi$ below 1.09 GeV }
\vspace*{-0.25cm}
\nin
%%%%%%%%%%%%%%%%%%%%%%%%%%%%%%%%%%%%%%%
%\vspace*{-0.3cm}
\subsection*{\b \bf The Born and unitarized  \boldmath$S$-wave amplitudes}
\begin{figure}
\begin{picture}(240,160)
  \SetWidth{0.6}
 % \Text(2,120)[l]{$T^{u}_{\pi}=$}
  \Text(20,145)[r]{$\gamma$}
  \Text(85,145)[l]{$\pi$}
  \Photon(25,145)(55,120){2}{4}
  \Photon(25,85)(55,100){2}{4}
  \Text(20,85)[r]{$\gamma$}
  \Text(85,85)[l]{$\pi$}
  \Line(55,120)(55,100)
  \Text(50,110)[r]{$\pi,V,A$}
  \Line(55,120)(80,145)
  \Line(55,100)(80,85)
  \Text(95,110)[r]{+}
  \Photon(105,145)(135,120){2}{4}
  \Photon(105,85)(135,100){2}{4}
  \Line(135,120)(135,100)
  \Text(130,110)[r]{$\pi,V,A$}
  \Line(135,120)(160,97)
  \Line(135,100)(160,130)
  \Text(175,110)[r]{+}
  \Photon(170,145)(200,110){2}{4}
  \Photon(170,85)(200,110){2}{4}
  \Line(200,110)(220,135)
  \Line(200,110)(220,90)
%  \Text(20,40)[l]{+}
  \Photon(25,75)(55,50){2}{4}
  \Photon(25,5)(55,20){2}{4}
  \Line(55,50)(55,20)
  \Text(50,35)[r]{$+~~~P,V,A$}
  \Line(55,50)(75,35)
  \Line(55,20)(75,35)
  \Text(65,45)[b]{$P$}
  \Text(65,25)[t]{$\bar{P}$}
  \GCirc(75,35){4}{0.5}
  \Line(78,38)(99,50)
  \Line(78,32)(99,20)
  \Text(105,35)[r]{$~~$+}
  \Photon(110,75)(140,40){2}{4}
  \Photon(110,5)(140,40){2}{4}
  \Curve{(140,40)(160,55)(180,40)}
  \Curve{(140,40)(160,25)(180,40)}
  \Text(160,60)[b]{$P$}
  \Text(160,20)[t]{$\bar{P}$}
  \GCirc(180,40){4}{0.5}
  \Line(183,43)(195,60)
  \Line(183,37)(195,20)
%  \Text(204,60)[l]{$(P=\pi,K)$}
%  \Text(204,45)[l]{R=V($1^{--}$),}
%  \Text(204,30)[l]{\ \ \ \ \ A($1^{++}$),}
%  \Text(204,15)[l]{\ \ \ \ \ B($1^{+-}$)}
\end{picture}
\caption{Born and Unitarized amplitudes $T^{u}_P$: $P\equiv
\pi,~K;~V\equiv \rho,~\omega;~A\equiv b_1, ~h_1,~a_1$.   }
\label{unitarized} \label{fig:unitarized}
\end{figure}
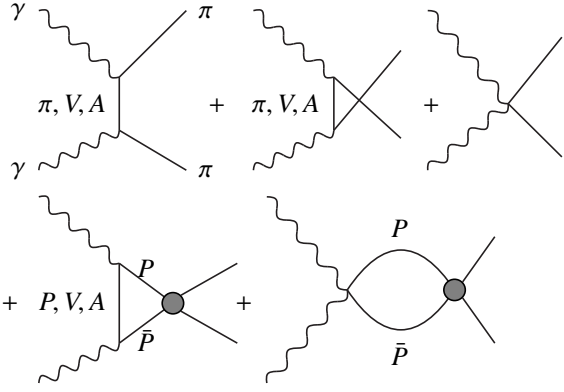

\nin 
%%%%%%%%%%%%%%%%%%%%%%%%%%%%
The Born and unitarized terms can be calculated unambiguously using the effective lagrangians. 
Taking into account the t-channel exchange of pion, vector and axial-vector mesons discussed 
in the previous section and shown in Fig. \ref{fig:unitarized}, the Born and unitarized parts of the amplitude given in Eq. (\ref{elmamp})
generalize to (normalized to $\alpha$):
\begin{eqnarray}
\left(
  \begin{array}{c}
    T_{\pi}^u \\
    T_{K}^u \\
  \end{array}
\right)&=&b_{\pi}\left(
                         \begin{array}{cc}
                           f^B_\pi+ &\tilde{f}^B_\pi\tilde{T}_{\pi\pi} \\
                           \ &\tilde{f}^B_\pi\tilde{T}_{\pi K} \\
                         \end{array}
                       \right)
                      % \nnb\\
                     %  &&
                       +b_K \left(
                      \begin{array}{cc}
                        \ &\tilde{f}^B_K\tilde{T}_{K\pi} \\
                        f^B_K+&\tilde{f}^B_K\tilde{T}_{KK} \\
                      \end{array}
                    \right)\nnb\\
                       &&+\sum_{V=\rho,\omega}b_V h_V^2\left(
                                                                    \begin{array}{cc}
                                                                      f^{BG}_V+&\tilde{f}^B_V\tilde{T}_{\pi\pi} \\
                                                                      \ &\tilde{f}^B_V\tilde{T}_{\pi K} \\
                                                                    \end{array}
                                                                  \right)\nonumber\\
&&+\sum_{V=K*^{+},K*^{0}}b_V h_V^2\left(
                                                             \begin{array}{cc}
                                                               \ &\tilde{f}^B_V\tilde{T}_{K\pi} \\
                                                               f^{BG}_V+&\tilde{f}^B_V\tilde{T}_{KK} \\
                                                             \end{array}
                                                           \right)\nonumber\\
&&+\sum_{A=a_1,b_1,h_1}b_A h_A^2\left(
                                                                    \begin{array}{cc}
                                                                      f^{BG}_A+&\tilde{f}_0^B(A)\tilde{T}_{\pi\pi} \\
                                                                      \ &\tilde{f}^B_A\tilde{T}_{\pi K} \\
                                                                    \end{array}
                                                                  \right)\
                                                                  ,
                                                                  \nnb\\
\end{eqnarray}

where the values of $h_{V,A}$ are in Eq. (\ref{eq:radcoupling}), $b_{\pi,\rho,\cdots}$ are Clebsh-Gordan coefficients for
projecting $\pi,\rho,...$ exchanges on the $I=0$ s-channel
amplitudes: 
\begin{eqnarray}
b_{\pi}&=&\sqrt{2}/\sqrt{3},\ \ \ \ b_{\rho}=\sqrt{3}/\sqrt{2},\ \ \
\
b_{\omega}=1/\sqrt{6},\nonumber\\
b_{K}&=&1/\sqrt{2},\ \ \ \
b_{K^{*+}}=b_{K^{*0}}=1/\sqrt{2},\nonumber\\
b_{a_1}&=&\sqrt{2}/\sqrt{3},\ \ \ \ b_{b_1}=\sqrt{3}/\sqrt{2},\ \ \
\ b_{h_1}=1/\sqrt{6}\ .
\end{eqnarray}
The partial $S$-wave Born amplitudes read :
\begin{eqnarray}
f^B_P&=&2L_1~:~P\equiv\pi,~K~;~~
f^{BG}_V=\frac{s}{8}[1-L_2]\ ,\nonumber\\
f^{BG}_{a_1}&=&\frac{s}{8}[-1+L_2]\
;~~~
f^{BG}_{b_1}=\frac{s}{8}[-1+L_2]\ ,
\end{eqnarray}
where $L_1(s,m_P^2)$ and $L_2(s,m_\pi, M_{V,A}^2)$ are functions analytic in the left hand
cut plane whose expressions are given in Appendix C of \cite{MENES}. The reduced amplitudes
are defined from the amplitudes in \cite{MNO,MNW} as:
\begin{eqnarray}
\tilde{T}_{\pi\pi}&\equiv&{G\over {\cal D}_P}\equiv {{T}_{\pi\pi}\over f_{\pi a}^2}=g_{\pi a}^2 P_{aa}+2g_{\pi a}g_{\pi b}P_{ab}+g_{\pi b}^2 P_{bb}\ ,\nonumber\\
\tilde{T}_{K\pi}&=&g_{\pi a}g_{Ka}P_{aa}+(g_{\pi a}g_{Kb}+g_{Ka}g_{\pi
b})P_{ab}+g_{\pi b}g_{Kb}P_{bb} \nonumber\\
&=&\tilde{T}_{\pi K}~,
\end{eqnarray}
%%%%%%%%%%%%%%%%%%%%%%%%%%%%%%%%%%%%%%%%%%%%%%%%%%%%%%%%%%%%%%%%%%%%%%%%%%%%%%%%%%%%%%%%%%%%%%%%%%%%%%%%%
%%%%%%%%%%%%%%%%%%%%%%%%%%%%%%%%%%%%%%%%%%%%%%%%%%%
where $f_{\pi a}$ is the shape function assumed to be the same for $\pi$ and $K$. $\tilde f^B_P~(P\equiv \pi, K)$ is defined in Eq. (\ref{elmamp}) like $\tilde f_P$ in Eq. (\ref{eq:ftilde}) expressed in terms of the functions $h_P(s)-h_P(0)$ but with $\tilde L_1$ replaced by $-\tilde L_1^2$ everywhere. For vector meson exchanges, the triangle loop function is
generalized to:
\begin{equation}
\tilde{f}_V^B(s)=\frac{1}{8\pi}[h_V^B(s)-h_V^B(0)]\,
\end{equation}
where
\begin{eqnarray}
h_V^B(s)&=&f_P(s)\Bigg{[}2M_V^2\tilde{L}_2(s)-\frac{2s(s-4m_P^2)}{m_P^2}\tilde{L}_1(s)\nonumber\\
&&+s\Big{[}\frac{2m_P^2-M_V^2}{M_V^2}-\ga \frac{M_V^2}{M_V^2-m_P^2}\dr^2\ln\frac{M_V^2}{m_P^2}\Big{]}\Bigg{]}\nonumber\\
&&-\frac{\sigma_{N0}}{s-\sigma_D}\Bigg{[}2M_V^2\tilde{L}_2(\sigma_D)\nnb\\
&&-\frac{2\sigma_D(\sigma_D-4m_P^2)}{m_P^2}\tilde{L}_1(\sigma_D)\nonumber\\
&&+\sigma_D\Big{[}\frac{2m_P^2-M_V^2}{M_V^2}-\ga\frac{M_V^2}{M_V^2-m_P^2}\dr^2\ln\frac{M_V^2}{m_P^2}\Big{]}\Bigg{]},
\end{eqnarray}
and where the functions $\tilde{L}_1(s)$ and $\tilde{L}_2(s)$ have been defined in the appendix C of \cite{MENES}. \\ For the axial-vector mesons, $h_A^B(s)$ is opposite in sign with the one for vector mesons given above. 
%%%%%%%%%%%%%%%%%%%%%%%%%%%%%%%%%%%
%\vspace*{-0.3cm}
\subsection*{\b \bf The Born and unitarized  \boldmath$D$-wave amplitudes}
\begin{figure}[h]%
\begin{center}%
%\vspace{2cm}
\includegraphics[width=6cm]{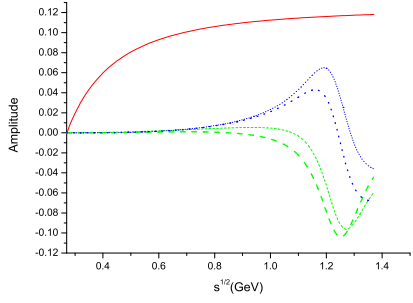}
\vspace*{-0.3cm}
%\vspace{-2cm}
\caption{\footnotesize  D-wave amplitudes in the two different models.   Born term (continuous - red line). 
Original model \cite{MENES}: real part  (dashed line - green); imaginary part (dotted line - blue ). New model: the same as for the original model but with thick lines. }
\label{fig:Dwave-model}
\end{center}%
\end{figure}

\nin 
%%%%%%%%%%%%%%%%%%%%%%%%%%%%%%%%%%%
Here we shall discuss two models, namely the one originally discussed in \cite{MENES} 
and a new model with a shape function inspired from the $S$-wave channel and which is expected 
to have a much better analytic property (absence of a pole in the left cut). We shall assume that the $D$-wave contribution is dominated by the helicity two $I=0$ amplitude which will be verified (a posteriori) from a fit analysis. \\
-- {\it The original model}  has been discussed in details in \cite{MENES}. 
For one pion exchange, the isospin $I=0$ 
$\gamma\gamma\rightarrow\pi\pi$ amplitude is:
\begin{equation}\label{old}
T_{\pi}^u=\sqrt{\frac{2}{3}}\Bigg{[} f_{\pi}^B+g_{\pi}^2\tilde{f}_{\pi}^B\ga\frac{s-4m_{\pi}^2}{s-s_0}\dr\mathcal
{D}^{-1}_{J=2}\Bigg{]}
\end{equation}
where $f_{\pi}^B$ is the Born term with helicity 2:
\begin{equation}
f_{\pi}^B=\frac{3}{8}\Bigg{[}\frac{1}{3}-\frac{2m_{\pi}^2}{s-4m_{\pi}^2}+\frac{8m_{\pi}^2}{s-4m_{\pi}^2}L_1(s)\Bigg{]}\
,
\end{equation}
and the unitary triangle one loop function $\tilde{f}_{\pi}^B$ satisfies:
\begin{equation}
\mathrm{Im}\tilde{f}_P^B=\theta(s-4m_P^2)\rho_P
f_P^B\frac{s-4m_P^2}{s-s_0}\ .
\end{equation}
The real part of $\tilde{f}_{\pi}^B$ can be derived from dispersion
relation with subtraction at $s=0$: 
\bea
\tilde{f}_{\pi}^B&=&\frac{3}{4}\Bigg{[}-\frac{s-4m_{\pi}^2}{s-s_0}\ga 2-\frac{s-4m_{\pi}^2}{3m_{\pi}^2}\dr\tilde{L}_1
-4\frac{s-4m_{\pi}^2}{s-s_0}\tilde{L}_1^2\nnb\\
&&+\frac{A_2
m_{\pi}^2}{s-s_0}+B_2\Bigg{]}\ ,
\eea
where
\begin{eqnarray}\label{AB}
A_2&=&\frac{s_0-4m_{\pi}^2}{m_{\pi}^2}\tilde{L}_1(s_0)\Bigg{[}\ga2-\frac{s_0-4m_{\pi}^2}{3m_{\pi}^2}\dr+4\tilde{L}_1(s_0)\Bigg{]}\
,\nonumber\\
B_2&=&\frac{m_{\pi}^2}{s_0}\ga \frac{13}{3}+A_2\dr\ .
\end{eqnarray}
-- {\it The new model} is introduced to avoid the left hand pole at $s=s_0$ of the old model. In this case the amplitude reads:
\begin{equation}\label{new}
T_{\pi}^u=\sqrt{\frac{2}{3}}\Bigg{[}f_{\pi}^B+g_{\pi}^2\tilde{f}_{\pi}^B|_{new}(s-4m_{\pi}^2)\mathcal
{D}^{-1}_{J=2}\Bigg{]}~,
\end{equation}
where the 2nd term is the unitarized amplitude. $\tilde{f}_{\pi}^B|_{new}$ is the triangle loop function due to one pion exchange:
\begin{equation}
\mathrm{Im}\tilde{f}_P^B|_{new}=\theta(s-4m_P^2)\rho_P
f_P^B\frac{s-4m_P^2}{{(s-s_0)^2}}\ .
\end{equation}
The real part of $\tilde{f}_{\pi}^B|_{new}$ can be derived from dispersion
relation with subtraction at $s=0$: 
\begin{eqnarray}
\tilde{f}_{\pi}^B|_{new}&=&\frac{3}{4}\Bigg{[}-\frac{s-4m_{\pi}^2}{{(s-s_0)^2}}\ga 2-\frac{s-4m_{\pi}^2}{3m_{\pi}^2}\dr\tilde{L}_1\nnb\\
&&-4\frac{s-4m_{\pi}^2}{{(s-s_0)^2}}\tilde{L}_1^2+\frac{A_2
m_{\pi}^2}{{(s-s_0)^2}}+\frac{B'_2}{{s-s_0}}\nnb\\
&&-\ga\frac{13}{3}\frac{m_{\pi}^2}{s_0^2}+\frac{A_2}{s_0^2}+\frac{B'_2}{-s_0}\dr\Bigg{]}~,
\end{eqnarray}
with:
\begin{equation}
B'_2=\lim_{s\rightarrow
s_0}\frac{d}{ds}\Bigg{[}(s-4m_{\pi}^2)\ga 2-\frac{s-4m_{\pi}^2}{3m_{\pi}^2})\tilde{L}_1
-4(s-4m_{\pi}^2\dr\tilde{L}_1^2\Bigg{]}
\end{equation}
%\begin{equation}
%f_{\pi}^u=g_{\pi}^2\tilde{f}_{\pi}^B(s-4m_{\pi}^2)\mathcal
%{D}^{-1}_{J=2}
%\end{equation}
We compare these two models in Fig. \ref{fig:Dwave-model}, where we can notice that the two models lead (almost) to the same amplitudes. From this figure, it is interesting to notice that there is a strong cancellation between the Born term and the real part of the unitarized amplitude around the $f_2(1270)$ pole, while the imaginary part of the amplitude is relatively small. This feature demonstrates that the $\gamma\gamma$ total width of the $f_2$ is dominated by its direct coupling as expected. 
%%%%%%%%%%%%%%%%%%%%%%%%%%%%
\nin
%%%%%%%%%%%%%%%%%%%%%%%%%%%%
In the following, we shall use the old model for our fitting procedure due only to a chronological procedure of our analysis. 
%%%%%%%%%%%%%%%%%%%%%%%%%%%%
%\vspace*{-0.3cm}
\subsection*{\b \bf The direct resonance couplings}
\nin 
%%%%%%%%%%%%%%%%%%%%%%%%%%%%
On the contrary the direct couplings of the resonances are model dependent where the polynomial ${\cal P}$ reflects the ambiguity from the dispersion relations\,\footnote{In \cite{MOUSSALAM2} these polynomial (subtraction constants) are related to the pion polarizabilities.}. We parametrize this contribution by introducing the effective photon-photon-resonance couplings for the $S$-waves \cite{MENES}:
\begin{equation}
T^{S}_{\pi}\equiv \alpha{\cal P\over {\cal D}_P}=\alpha{\sqrt{2}s\over {\cal D}_P}
\Big{[}(f_{\sigma\gamma}+sf'_{\sigma\gamma})\tilde{T}_{\sigma\pi}+(f_{f_0\gamma}+sf'_{f_0\gamma})\tilde{T}_{f_0
\pi}\Big{]},
\end{equation}
where the reduced amplitudes are:
\begin{eqnarray}
\tilde{T}_{\pi\sigma}&=&[g_{\pi a}(s_{Rb}-s)-g_{Kb}(g_{\pi a}g_{Kb}-g_{Ka}g_{\pi b})\tilde{f}_K]
%/\mathcal {D}
\ ,\nonumber\\
\tilde{T}_{\pi f_0}&=&[g_{\pi b}(s_{Ra}-s)-g_{Ka}(g_{\pi
b}g_{Ka}-g_{Kb}g_{\pi a})\tilde{f}_K]~.
%/\mathcal {D}_P
\nnb\\
\end{eqnarray}
The functions $f_0(s)$, $\tilde{f}_P(s)(P=\pi,K)$, $\mathcal {D}_P(s)$
and bare parameters $\sigma_D$, $s_A$, $s_{Ra}$, $g_{\pi a}$,
$g_{Ka}$, $s_{Rb}$, $g_{\pi b}$, $g_{Kb}$ are defined
in~\cite{MNO,MNW} and given in Tables 2 and 3 of ref. \cite{MNW}.
Similarly, we have for the $D$-waves \cite{MENES}:
\beq
T^{D}_{\pi}=\ga{\alpha\over\sqrt{2}}\dr \Big{[}s^2 f^{\lambda=0}_{f_2\gamma}+
sf^{\lambda=2}_{f_2\gamma}\Big{]}\tilde T_{\pi f_2}~,
\label{eq:f2coupling}
\eeq
with the normalization:
\beq
\Gamma_{f_2\to\gamma\gamma}={4\over 3}\ga \alpha  f_{2\gamma}^{\lambda=2}\dr^2M_{f2}^3~,
\eeq
if one assumes that the $\lambda=0$ helicity contribution is negligible\,\footnote{If we let free the two couplings of the $\lambda=0$ and 2 components of the $f_2(1270)$ in the fit, we find that the $\lambda=0$ coupling is negligible confirming our assumption. We also notice that at the $f_2$-pole, there is a strong cancellation between the Born and rescattering contributions (see also \cite{MENES})
which justifies the identification of the $f_2\to\gamma\gamma$ total width given by PDG \cite{PDG} to the ``direct" width.}. Using the PDG value $(2.6\pm 0.24)$ keV \cite{PDG} for the $\gamma\gamma$ width,
we deduce:
\beq
|f_{2\gamma}^{\lambda=2}|=0.136~{\rm GeV}^{-1}~.
\eeq
%Notice that in doing the fits of the $S$- and $D$-waves data, the ${\cal D}_P$  function has a zero on the negative real axis, such that both the direct and unitarized amplitudes have a pole on the left hand cut in addtion to the normal one in the right hand cut. In the strict analytical final state interaction approach, which is assumed to build the right hand cut singularities while the left ones are exactly known, one should remove this pole. However, within our approximation, we assume that this pole takes into account part of the $\pi\pi$ and $\bar KK$ interactions, which would have been neglected.
%%%%%%%%%%%%%%%%%%%%%%
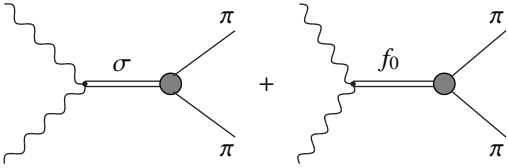
\begin{figure}
\begin{picture}(240,70)
%  \Text(10,40)[r]{$T^{D}_{\pi}=$}
  \Photon(20,65)(50,35){2}{4}
  \Photon(20,5)(50,35){2}{4}
  \Vertex(50,35){1}
  \Line(50,36)(79,36)
  \Line(50,34)(79,34)
  \Text(64,40)[b]{$\sigma$}
  \GCirc(82,35){4}{0.5}
  \Line(83,38)(106,55)
  \Line(83,32)(106,15)
  \Text(106,60)[r]{$\pi$}
  \Text(106,10)[r]{$\pi$}
  \Text(115,35)[l]{+}
  \Photon(130,65)(150,35){2}{4}
  \Photon(130,5)(150,35){2}{4}
  \Line(150,36)(180,36)
  \Line(150,34)(180,34)
  \Text(164,40)[b]{$f_0$}
  \Vertex(150,35){1}
  \GCirc(184,35){4}{0.5}
  \Line(187,38)(207,55)
  \Line(187,32)(207,15)
  \Text(207,60)[r]{$\pi$}
  \Text(207,10)[r]{$\pi$}
\end{picture}
\caption{Direct couplings of the resonances to $\gamma\gamma$}
\label{direct}
\end{figure}
%%%%%%%%%%%%%%%%%%%%%%%%%%%%%%%%%%%%%%%%%%%%%%%%%%%

%%%%%%%%%%%%%%%%%%%%%%%%%%%%
%\vspace*{-0.3cm}
\section{Fitting $\gamma\gamma\to\pi^0\pi^0$ just above the $\bar KK$ threshold}
%\vspace*{-0.25cm}
%\nin 
%%%%%%%%%%%%%%%%%%%%%%%%%%%%
\subsection*{\b \bf Fitting procedure}
\begin{figure}[hbt]
\centerline{\includegraphics[width=7.cm]{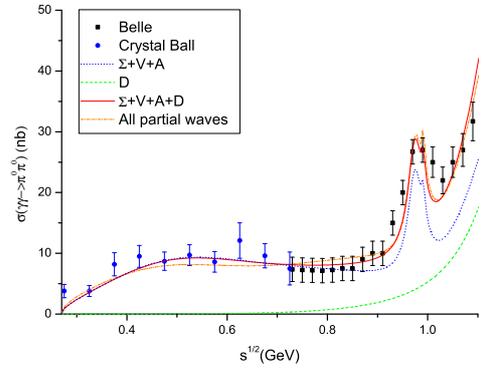}}
\centerline{\includegraphics[width=7.cm]{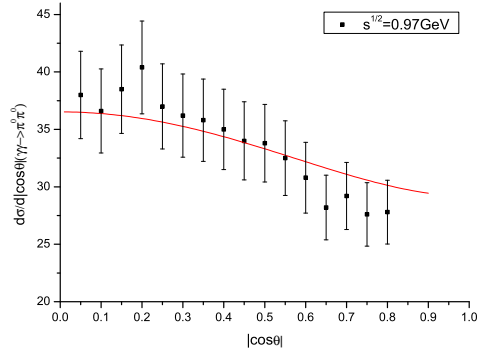}}
%{\epsfig{figure=mpsi2mc.eps,height=70mm}}
\caption{\scriptsize a) Fit of the $\gamma\gamma\to \pi^0\pi^0$ total cross-section ($\sqrt{s}\leq1.09~\mathrm{GeV}$ and $|\cos\theta|\leq 0.8$)   for the Set 3 of hadronic parameters \cite{MNW}, where
$\chi^2/{ndf}={39.5}/{41}=0.96$ is minimal. The data come from Crystal Ball (blue full circle) and from Belle (black full square); dotted blue (S-channel contribution); dashed green (D-wave contribution); continuous red (total contribution); dotted-dashed salmon (sum of all partial waves); b) Comparison of the effects of the sum of all partial waves (dashed-dotted salmon) with sum of S+D waves (continuous red) and the Belle data on the $\gamma\gamma\to \pi^0\pi^0$ differential cross-section at $\sqrt{s}=0.97$ GeV. }
 \label{fig:fit1.09}
\end{figure}

\nin 
%%%%%%%%%%%%%%%%%%%%%%%%%%%%
In so doing, we fix the value of the $\lambda=2$ components of the $f_2(1270)$ using the $f_2~\gamma\gamma$ width given by PDG and neglect the $\lambda=0$ one. We use as inputs the Set 2 and Set 3 hadronic parameters obtained in \cite{MNW} using the largest range of hadronic data. 
Then, we perform a fit of the 
four direct couplings from the $\gamma\gamma\to\pi^0\pi^0$ total  cross-section up to $\sqrt{s}\approx 1$ GeV\,\footnote{We choose not to fit much above 1 GeV in order to avoid the possible contribution of an eventual $f_0(1370)$ and to minimize the contribution of the $D$-wave.} . We move $\sqrt{s}$ around 1 GeV and looks for the minimum $\chi^2/ndf$ for the total cross-section,
which is obtained at $\sqrt{s}$= 1.09 GeV, where $\chi^2/ndf=39.5/41=0.96$ for e.g. Set 3 of the hadronic parameters, which we show in Fig. \ref{fig:fit1.09}a). In Fig. \ref{fig:fit1.09}b), we compare the $\gamma\gamma\to\pi^0\pi^0$ differential cross-section at $\sqrt{s}=0.97$ GeV for sum of all partial  waves with the sum of the S+D waves and with the Belle data. 
%%%%%%%%%%%%%%%%%%%%%%%%%%%%%%%%
%%%%%%%%%%%%%%%%%%%%%%%%%%%%%%%
The results of the fit are given in Tables \ref{tab:bare} and  \ref{tab:physical}. %%%%%%%%%%%%%%%%%%%%%%%%%%%
%\vspace*{-0.5cm}
{\footnotesize
\begin{table}[hbt]
\setlength{\tabcolsep}{.8pc}
 \caption{\scriptsize   Fitted values of the bare couplings of $\sigma$ and $f_0(980)$ in units of GeV$^{-1}$ for $(f_{S\gamma})$ and GeV$^{-2}$ for $(f'_{S\gamma})$ using Set 2 and Set 3 of the hadronic parameters from \cite{MNW}. The $\gamma\gamma\to\pi\pi$ total cross-section has been fitted until  $\sqrt{s}$=1.09 GeV (minimum value of $\chi^2/ndf$) and the differential cross-section until $\sqrt{s}=0.97$ GeV.}
 %the $\lambda=0$ over the $\lambda=2$ branching fraction of the $f_2: ~r_{0/2}\equiv 
% {\Gamma^{\lambda=0}/ \Gamma^{\lambda=0+2}}$}
   {\footnotesize
\begin{tabular}{llllll}
&\\
\hline
%\multicolumn{5}{l}{Bare couplings in units of GeV$^{-1}~(f_{S\gamma})$ and GeV$^{-2}~ (f'_{S\gamma})$}\\
%\hline
%\\
&$f_{\sigma\gamma}$&$f'_{\sigma\gamma}$&$f_{f_0\gamma}$ &$f'_{f_0\gamma}$&$\chi^2/ndf$\\
%\\
\hline
\\
Set 2&2.68&-2.70&0.85&-1.18&51.4/41=1.25\\
%&\\
Set 3&3.03&-3.07&1.03&-1.43&39.5/41=0.96\\
&\\
\hline
%\multicolumn{7}{l}{Physical couplings in units of $\alpha\times 10^{-3}$ GeV}\\
%\hline
%&\multicolumn{1}{l}{$g^{dir}_{\sigma\gamma}$}&\multicolumn{1}{l}{$g^{resc}_{\sigma\gamma}$}&\multicolumn{1}{l}{$g^{tot}_{\sigma\gamma}$}&\multicolumn{1}{l}{$g^{dir}_{f_0\gamma}$}&\multicolumn{1}{l}{$g^{resc}_{f_0\gamma}$}&\multicolumn{1}{l}{$g^{tot}_{f_0\gamma}$}\\
%\hline
%One res.\cite{MNO}&\multicolumn{1}{l}{7+{\rm i}31}&\multicolumn{1}{l}{90+{\rm i}116}&\\
%Set 2&\multicolumn{1}{l}{9+{\rm i}37}&\multicolumn{1}{l}{42+{\rm i}109}&\multicolumn{1}{l}{51+{\rm i}146}&\multicolumn{1}{l}{-73-{\rm i}12}&\multicolumn{1}{l}{131+{\rm i}14}&\multicolumn{1}{l}{57+{\rm i}2}\\
%Set 3&\multicolumn{1}{l}{5+{\rm i}42}&\multicolumn{1}{l}{33+{\rm i}110}&\multicolumn{1}{l}{38+{\rm i}152}&\multicolumn{1}{l}{-71-{\rm i}10}&\multicolumn{1}{l}{124+{\rm i}15}&\multicolumn{1}{l}{53+{\rm i}5}\\
%\hline
%\\
%\hline
\end{tabular}
}
\label{tab:bare}
\end{table}
}
%%%%%%%%%%%%%%%%%%%%%%%%%%%%%%%%%%%%%%%%%%%%%%%%%%%
\begin{table}[hbt]
\begin{center}
\footnotesize\setlength{\tabcolsep}{0.15pc} \caption{The same fitting procedure as in Table \ref{tab:bare}
but for  the physical couplings (in units of $\alpha\times 10^{-3}$ GeV) and for the two photon widths (in units of keV) of the $\sigma$ and $f_0(980)$.  Each contributions  of different mesons exchanged in the t-channel and in the loops are shown explicitly. $D$ corresponds to the $D$-wave contribution.
%  using Set 2 and Set 3 of the hadronic parameters \cite{MNW}. The $\gamma\gamma\to\pi\pi$ total cross-section has been fitted until  $\sqrt{s}$=1.09 GeV (minimum value of $\chi^2/ndf$) and the differential cross-section until $\sqrt{s}=0.97$ GeV. The hadronic
%parameters are taken from Set 2 and Set 3 of \cite{MNW}.
}
\vspace*{0.25cm}
\begin{tabular}{lccccccc}\hline
 Fit& $\pi$ &{\it $\Sigma\equiv\pi$+K} & {\it $\Sigma$+V} &{\it $\Sigma$+A} & {\it $\Sigma$+V+A} & {\it $\Sigma$+D} & {\it $\Sigma$+V+A+D} \\
\hline 
&\\
Set 2\\
$g^{dir}_{\sigma}$ & -2+49i & 19+15i & 26+19i & 10+22 i & 20+27i & 8+24i & 9+37i \\
$g^{resc}_{\sigma}$ & 65+99i & 66+103i & -23+185i & 131+28i & 42+109i & 66+103i & 42+109i \\
$g^{tot}_{\sigma}$ & 63+148i & 85+118i & 3+204i & 141+50i & 62+136i & 74+127i & 51+146i\\
\  & \ & \ & \  & \ & \ & \ & \ \\
$\Gamma^{dir}_{\sigma}$  & 0.27 & 0.06 & 0.11 & 0.07  & 0.13 & 0.07 & 0.16 \\
$\Gamma^{resc}_{\sigma}$ & 1.60 & 1.71 & 3.97 & 2.05  & 1.53 & 1.71 & 1.53 \\
$\Gamma^{tot}_{\sigma}$  & 2.90 & 2.36 & 4.66 & 2.51  & 2.50 & 2.43
& 2.67 \\
\ & \ & \ & \ & \ &\ & \ & \ \\
$g^{dir}_{f_0}$  & -125+3i  & -4-5i & -109-18i & 26-6i & -75-14i & -5-4i & -73-12i \\
$g^{resc}_{f_0}$ & 12+12i   & 95+16i & 47+16i & 178+14i & 130+14i & 95+16i & 131+14i \\
$g^{tot}_{f_0}$  & -113+15i & 91+11i & -62-2i & 204+8i & 55 & 90+12i & 52+2i \\
\ & \ & \ & \ & \ & \ & \ & \ \\
$\Gamma^{dir}_{f_0}$ & 0.82 & 0.002 & 0.63 & 0.04 & 0.31 & 0.002 &0.29 \\
$\Gamma^{resc}_{f_0}$ & 0.02 & 0.48 & 0.13 & 1.67 & 0.90 & 0.48 & 0.90 \\
$\Gamma^{tot}_{f_0}$ & 0.68 & 0.43 & 0.20 & 2.19 & 0.16 & 0.42 &0.17 \\ 
&\\
\hline
&\\
Set 3\\
$g^{dir}_{\sigma}$ & 18+29i & 12+28i & 27+16i & 0.2+42i & 17+33i & 0.3+36i  & 5+42i \\
$g^{resc}_{\sigma}$ & 58+96i & 58+100i & -41+191i & 132+20i& 33+109i & 58+100i & 33+109i \\
$g^{tot}_{\sigma}$ & 76+125i & 70+128i & -14+207i & 132+62i & 50+142i & 58+136i & 37+152i \\
\  & \ & \ & \  & \ & \ \\
$\Gamma^{dir}_{\sigma}$  & 0.13 & 0.10 & 0.11 & 0.19  & 0.15 & 0.14 & 0.20 \\
$\Gamma^{resc}_{\sigma}$ & 1.36 & 1.44 & 4.22 & 1.94  & 1.43 & 1.44 & 1.43 \\
$\Gamma^{tot}_{\sigma}$  & 2.34 & 2.35 & 4.70 & 2.32  & 2.49 & 2.43
& 2.67 \\ 
\ & \ & \ & \ & \ &\ & \ & \ \\
$g^{dir}_{f_0}$ & -99+42i & -10-8i & -95-14i & 22-8i & -65-11i & -13-7i & -71-10i \\
$g^{resc}_{f_0}$ & 12+8i & 96+6i & 51+7i & 170+3i & 124+4i & 96+6i & 124+15i \\
$g^{tot}_{f_0}$ & -87+50i & 86-2i & -44-7i & 192-5i & 59-7i & 83-i& 54+5i \\
\ & \ & \ & \ & \ & \ & \ & \ \\
$\Gamma^{dir}_{f_0}$ & 0.61 & 0.01 & 0.48 & 0.03 & 0.23 & 0.01 & 0.27 \\
$\Gamma^{resc}_{f_0}$ & 0.01 & 0.49 & 0.14 & 1.51 & 0.81 & 0.49 & 0.81 \\
$\Gamma^{tot}_{f_0}$ & 0.53 & 0.39 & 0.10 & 1.93 & 0.19 & 0.36 & 0.15 \\
&\\
 \hline
%\ & \ & \ & \ & \ &\ \\
%$\Gamma^{dir}_{f_0}$ & 0.004 & 0.24 & 0.01 & 0.28 \\
%$\Gamma^{resc}_{f_0}$ & 0.49 & 0.81 & 0.49 & 0.81 \\
%$\Gamma^{tot}_{f_0}$ & 0.44 & 0.17 & 0.34 & 0.14 \\
%\ & \ & \ & \ & \ \\
%$\chi^2_{d.o.f}$ & $\frac{43.8}{25}=1.75$ & $\frac{44.5}{25}=1.78$ &
%$\frac{40.3}{39}=0.82$  & $\frac{30.8}{39}=0.79$ \\ \hline
\end{tabular}
\label{tab:physical}
\end{center}
\end{table}
%%%%%%%%%%%%%%%%%%%%%%%%%%%%%%%%%%%%%%%%%%%%%%%%%%%%%%%
{\scriptsize
\begin{table}[hbt]
\setlength{\tabcolsep}{0.2pc}
 \caption{\scriptsize Summary of the fitted values of the $\gamma\gamma$ decay width(in unit of $\mathrm{keV}$ using Set 2 and Set 3 of hadronic parameters from \cite{MNW} and comparison with some other determinations and the PDG08 value.
 The total cross-section has been fitted until $\sqrt{s}=1.09$ GeV (minimum value of $\chi^2/ndf$) and the differential cross-section unitl $\sqrt{s}=0.97$ GeV. The best $\chi^2/ndf=0.96$ is obtained from Set 3.}
 \vspace*{0.25cm}
{\footnotesize
\begin{tabular}{lcccccccccc}\hline
&Set 2&Set 3&\cite{MNO} &\cite{OLLER08} &  \cite{PEN08} & \cite{PEN08}& \cite{ZHENG09}&\cite{PRADES08}&\cite{ACHASOV08}&PDG \cite{PDG} \\
\hline
&\\
$\sqrt{s}$&1.09&1.09&0.8&0.8&1.44&1.44&1.4&&1.44&\\
$r_{0/2}$ 
&0.0 & 0.0  &&&0.13&0.26&0.15 \\
%$\chi^2/ndf$&1.25&0.96\\
%\hline
&\\
$\Gamma^{dir}_{\sigma}$ & 0.16 & 0.20 & 0.13   \ & &&&&&0.010&\\ % 0.06
$\Gamma^{resc}_{\sigma}$ & 1.53 & 1.43 & 2.70 & \ & &&&&\\ % 2.4
$\Gamma^{tot}_{\sigma}$  & 2.67 & 2.67 &3.90 & 1.7 &  3.1 & 2.4& 2.1&1.2&\\ % 1.8
\ & \ & \ & \ & \ & \ & \ \\
$\Gamma^{dir}_{f_0}$  & 0.29 & 0.27 & \  &&&&&&0.015& \ \\
$\Gamma^{resc}_{f_0}$  &  0.90 & 0.81 &\ & &&&& \ \\
$\Gamma^{tot}_{f_0}$  & 0.17 & 0.15 & \ & \ & 0.42 & 0.10 &0.13&&&$0.29\pm 0.08$ \\
%\ & \ & \ & \ & \ & \ \\
%\chi^2_{d.o.f} & $\frac{23.3}{24}=0.97$ & $\frac{21.6}{24}=0.90$ & \ & \ & \ \\
\hline
\end{tabular}
} \label{tab:width}
\end{table}
}
%%%%%%%%%%%%%%%%%%%%%%%%%%%%%%%%%%%%%%%%%%%%%%%%
\subsection*{\b \bf Comments on the results}
% and possible nature of the $\sigma$ and $f_0(980)$}
\nin
%%%%%%%%%%%%%%%%%%%%%%%%%%%%%%%%%%
Our results are summarized in Tables \ref{tab:physical} and \ref{tab:width}:
\\
{\it-- The direct part of the $\gamma\gamma$ widths:} one can notice that the $K$ exchange and $K$-loop tend to decrease the direct width of the $\sigma$ meson which is compensated by the V+A contributions, such that at the end, the final result is compatible with the one 0.13 keV obtained in the case: 1 resonance $\oplus$ 1 channel obtained below 0.7 GeV \cite{MNO}. We consider as a final result the average from Set 2, Set 3 and from the one resonance $\oplus$ one channel analysis below 0.7 GeV of \cite{MNO} for the $\sigma$ and the average from  Set 2 and Set 3 for the $f_0(980)$ (see Table \ref{tab:width}):
 \bea
 \Gamma_\sigma^{dir}= 0.16(4)~{\rm keV},~~~ \Gamma_{f_0}^{dir}\simeq 0.28(1)~{\rm keV}~.
   \label{eq:resdir}
   \eea
%   where the value of $ \Gamma_\sigma^{dir}$ is in good agreement with the one in \cite{MNO} from $\gamma\gamma\to\pi\pi$ analysis below 0.7 GeV. 
For a consistent comparison with some other theoretical estimates (QSSR,...) obtained in the real axis,
we translate these widths from the residues to
   the one evaluated at
%    ~~\lrar ~~\Gamma_\sigma^{dir}\vert_{on-shell}= ~1.4(2)~{\rm keV}\nnb\\
%  \Gamma_{f_0}^{dir}&\simeq& \Gamma_{f_0}^{dir}\vert_{on-shell}\simeq 0.28(1)~{\rm keV},
 the on-shell $\sigma$ mass or Breit-Wigner mass defined in \cite{SIRLIN,MNO} when the amplitude is purely imaginary at the phase 90$^0$:
 \beq
 {\rm Re}~{\cal D}(M^2_\sigma)=0~~~~\lrar ~~~~M^{os}_\sigma\simeq 0.92 ~{\rm GeV}.
 \label{eq:onshell}
 \eeq
 In this way, we obtain:
 \beq
 \Gamma_\sigma^{dir}\vert_{on-shell}= ~1.2(3)~{\rm keV},~~~
\Gamma_{f_0}^{dir}\vert_{on-shell}\approx \Gamma_{f_0}^{dir} ,
  \label{eq:resdir2}
\eeq
which are similar with the results obtained by using a Breit-Wigner parametrization of the data \cite{MNO}. \\
 {\it -- The rescattering part} of   the $\gamma\gamma$ widths are  (in units of keV):
 \bea
 \Gamma_\sigma^{resc}= 1.89(81)~{\rm keV}~,~~~~~~~~  \Gamma_{f_0}^{resc}= 0.85(5) ~{\rm keV}~,
   \label{eq:resresc}
 \eea 
where we take the average from Set 2, Set 3 and from the one resonance $\oplus$ one channel analysis below 0.7 GeV of \cite{MNO} for the $\sigma$ and the average from  Set 2 and Set 3 for the $f_0(980)$ (see Table \ref{tab:width}).
 One can notice that in both cases, the rescatterings are relatively large indicating the important r\^ole of
 meson loop contributions in the $\gamma\gamma$ widths of the scalar mesons.The large couplings of scalar to meson loops are often interpreted in the current literature as being related to their four-quark or/and molecule structure. However, these large couplings to $\pi\pi$ and $\bar KK$  are also expected if the $\sigma,~f_0$ have large gluon component and violate OZI rule in their hadronic decays \cite{VENEZIA,SNG0,SN06}. 
 % In this sense, these contributions might be included into the so-called continuum contribution within the QCD spectral sum rules (QSSR) approach.}. \\
 In the case of the $\sigma$ meson, one can notice the large effect due to vector mesons which is partly compensated by the one of the axial-vector mesons. 
Compared with the result 2.7 keV from the analysis below 0.7 GeV with a pion loop \cite{MNO} , one can notice that $ \Gamma_\sigma^{resc}$ has been affected by the presence of the $f_0(980)$ when doing the fitting procedure.  
In the case of the $f_0(980)$, the effect of the $K$ and of $V+A$ are quite important
  \\
 {\it -- The total $\equiv$ direct + rescattering part} of the $\gamma\gamma$ widths are  (in units of keV):
 \bea
 \Gamma_\sigma^{tot}= 3.08(82)~{\rm keV}~,~~~~~~~~  \Gamma_{f_0}^{tot}= 0.16(1)~{\rm keV} ~.
   \label{eq:restot}
 \eea 
 %%%%%%%%%%%%%%%%%%%%%%%%%%%%
%\vspace*{-0.3cm}
\section{Comparison with some other results}
%\vspace*{-0.25cm}
%\nin 
%%%%%%%%%%%%%%%%%%%%%%%%%%%%
 %%%%%%%%%%%%%%%%%%%%%%%%%%%%%%%%%%%%%%%%
\subsection*{\b \bf Dispersion relations}
% and possible nature of the $\sigma$ and $f_0(980)$}
\nin
%%%%%%%%%%%%%%%%%%%%%%%%%%%%%%%%%%%%%%%%
A comparison of the total width in Eq. (\ref{eq:restot})  with the results obtained using dispersion relations \cite{OLLER08,PEN08,ZHENG09,PRADES08} \footnote{Some calculations based on the $\pi$ and $K$ loops can also be found in \cite{OLLER97}.} is shown in Table \ref{tab:width}. The results can only be compared for the total width, as the authors have not performed the (model-dependent) separation of the direct and rescattering processes done in this approach. Our results for the $\sigma$ are in better agreement with
 the ones from \cite{MNO,PEN08,ZHENG09}. The one of the $f_0$ is in between the results of \cite{PEN08,ZHENG09} and of the 
 PDG value  \cite{PDG}. However, the PDG value seems to be more consistent with our value of the direct width which might be identified with the one using a Breit-Wigner parametrization of the data.
 %%%%%%%%%%%%%%%%%%%%%%%%%%%%%%%%%%%%%%%%%%%%%%%%
\subsection*{\b \bf Model of \cite{ACHASOV08}}
% and possible nature of the $\sigma$ and $f_0(980)$}
\nin
%%%%%%%%%%%%%%%%%%%%%%%%%%%%%%%%%%%%%%%%%%%%%%%%
This approach of \cite{ACHASOV06,ACHASOV08,ACHASOV082} presents some similarities with ours where the direct and rescattering processes
can be also separated.  However, we differ in the treatment of the hadronic $\pi\pi\to\pi\pi/\bar KK$ process as an (a priori ad hoc) background phase has been explicitly introduced which multiplies the resonance contributions to the amplitude in this approach. 
%in \cite{ACHASOV06,ACHASOV08,ACHASOV082}. 
Besides this difference, the authors also have a different (philosophical) view on the problem, where they claim that considering the residues of the $\sigma$ pole for the hadronic amplitude does not give a clear understanding of the nature of the $\sigma$, while, in our paper, we, on the contrary, claim that the residues in the complex plane can clarify its nature.  For a better understanding of the quantitative
difference between the two approaches, we look in details into the analysis of \cite{ACHASOV08}:\\
-- We  note that the authors use the bare couplings
for predicting the $\gamma\gamma$ width of the resonances while we transform the bare couplings (which are real numbers) to the residues in the complex plane by analytically continuing these results to the 2nd sheet. In fact, we expect that working with the bare couplings for the wide complex $\sigma$ meson cannot (a priori) be a good approximation. \\
-- We also note that the authors only include the $S$ and $D$-waves and neglect the contributions due to the vector and axial-vector mesons (V+A) which is a bad approximation for the direct width predictions (compare the last two columns of Table \ref{tab:physical}) though the numerical fits of the total cros-section are similar in the two cases (see Fig. \ref{fig:fit1.09}) because of the dominant contribution of the rescattering amplitude which remains almost constant. \\
%However, though the numerical fits are similar (see Fig. \ref{fig:fit1.09}) with a slightly larger $\chi^2/ndf= 1.16$ for $\Sigma+D$-waves,  instead of 0.96 for $\Sigma+D+V+A$, the values of the widths are largely affected as on can deduce from the last two columns of Table \ref{tab:physical}. \\
-- From Table  \ref{tab:physical}, one can indeed deduce within the approximation of \cite{ACHASOV08}
by retaining the S+D wave contributions that the value of the $f_0(980)$ direct width is about (0.002-0.01) keV for Sets 2 and 3
which agrees with the tiny value obtained in \cite{ACHASOV08}. \\
-- For the $\sigma$, a comparison of our result with the one of \cite{ACHASOV08} is less direct due to the large width of the $\sigma$ and to the important effect of the background phase which multiplies the resonance hadronic amplitude. Indeed, a factorization of this phase can allow a 90$^0$ phase at a low on-shell mass of about 500 MeV, while it is about 1 GeV, similar to the one obtained from a Breit-Wigner paramterization, in our approach  [Eq. (\ref{eq:onshell})].
If one uses the previous inputs (bare couplings + $S$ and $D-$waves),  one can see from the value of the bare coupling obtained in \cite{MNO} that the direct width of the $\sigma$ would be about 0.02 keV for a $\sigma$ mass of 0.42 GeV which is similar to the result of \cite{ACHASOV08}.
% For the latter, we have assumed that the bare and physical couplings are about the same, which can be a good approximation because the $f_0(980)$ is narrow. 
The results of our tests agree with the ones of \cite{ACHASOV08} but these results might not be realistic due the drawbacks which we have mentioned above.
%%%%%%%%%%%%%%%%%%%%%%%%%%%%%%%%
\begin{figure}[hbt]
\centerline{\includegraphics[width=7.cm]{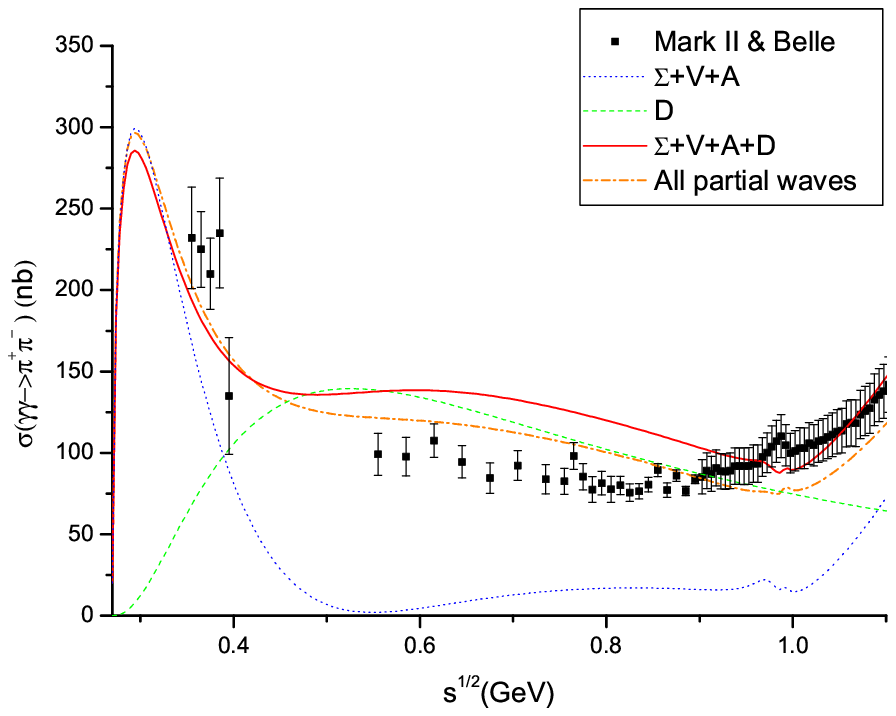}}
\centerline{\includegraphics[width=7.cm]{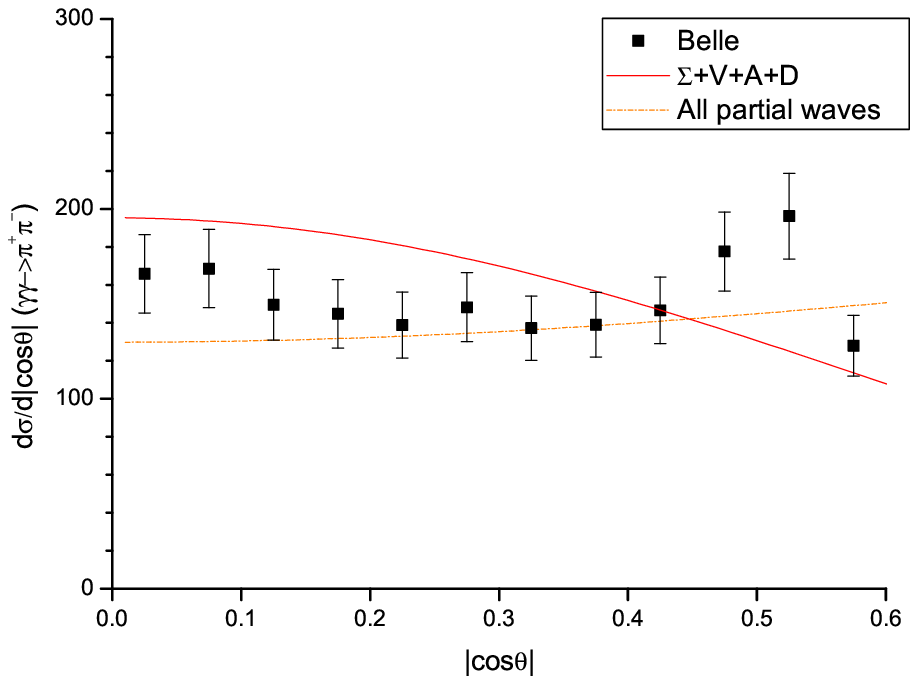}}
%{\epsfig{figure=mpsi2mc.eps,height=70mm}}
\caption{\scriptsize a) Predictions of the $\gamma\gamma\to \pi^+\pi^-$ total cross-section ($\sqrt{s}\leq1.09~\mathrm{GeV}$ and $|\cos\theta|\leq 0.6$) using the fitted parameters from $\gamma\gamma\to \pi^0\pi^0$. The data come from Mark II and Belle: dotted blue (S-channel contribution including direct couplings); dashed green (D-wave contribution); continuous red (S+D partial waves); dot-dashed salmon (sum of all partial waves); b) Comparison of the effects of the sum of all partial waves (dashed-dotted salmon) with sum of S+D waves (continuous red) and the Belle data on the differential $\gamma\gamma\to \pi^+\pi^-$ cross-section at $\sqrt{s}=0.95$ GeV. }
 \label{fig:pi+pi-}
\end{figure}
%%%%%%%%%%%%%%%%%%%%%%%%%%%%%%%
%%%%%%%%%%%%%%%%%%%%%%%%%%%%%%%%%%%%%%%%%%
\subsection*{\b \bf $\gamma\gamma\to \pi^+\pi^-/K^+K^-$ data}
% and possible nature of the $\sigma$ and $f_0(980)$}
\nin
%%%%%%%%%%%%%%%%%%%%%%%%%%%%%%%%%%%%%%%%%%
We use the previous fitted values of the parameters from $\gamma\gamma\to \pi^0\pi^0$ for predicting
the $\gamma\gamma\to \pi^+\pi^-$ process.  The results for the total and differential cross-sections are shown in Fig. \ref{fig:pi+pi-}, where one can notice that the prediction is not good between 0.5 to 0.9 GeV if one only retains the $S$ and $D$-waves in the partial waves of the Born term. A similar problem has been encountered by the authors of \cite{ACHASOV08} (Fig 4) who solve this (Fig. \ref{fig:pi+pi-}) by introducing a form factor (which looks ad hoc) from \cite{POPPE} for the $D$-wave Born contribution which is important in the charged channel \footnote{A similar observation can be done for the process 
$\gamma\gamma\to K^+K^-$, where the Born contributions due to the $S$ and $D$-waves are also very important and dominates over the scalar meson contributions, as can be shown in Fig. 10 of \cite{HARJES}, which we have checked.}. Instead, we improve  our analysis by adding all higher partial waves in the Born term and by including the direct terms of the $S$-waves. One can see in Fig. \ref{fig:pi+pi-} , that the predictions using the parameters from $\gamma\gamma\to\pi^0\pi^0$ are quite good compared with the data without introducing any ad hoc form factors.  
%%%%%%%%%%%%%%%%%%%%%%%%%%%%%%%%
\section{On some possible substructures of the $\sigma$ and $f_0(980)$}
\vspace*{-0.25cm}
\nin
%%%%%%%%%%%%%%%%%%%%%%%%%%%%%%%%%%
Our work has been motived for shedding light on the possible substructure of the $\sigma/f_0(600)$ and $f_0(980)$ by analyzing their $\gamma\gamma$ widths, where the value of direct width is of special interest for a comparison with theoretical calculations based on quark and/or gluon loops.  
%in addition to their hadronic couplings obtained in \cite{KMN,MNO}
\\
{\it -- The $\sigma$ meson}: in our previous analysis of the hadronic couplings \cite{MNW,KMN} from $\pi\pi\to\pi\pi,~\bar KK$ processes, we have noticed that because of the large size of its coupling to $\pi\pi$ and $\bar KK$ [$|g_{\sigma K^+K^-}|/|g_{\sigma\pi^+\pi^-}|=0.37(6)$] (see also \cite{MENES,OTHERS,ZHENG09}), the $\sigma$ cannot be mainly a  $\pi\pi$ molecule or/and a four-quark state components, while its broad width cannot be explaIned by a large $\bar qq$ component.  These observations go in line to a large gluon component  expected from a low-energy theorem analysis in \cite{VENEZIA,SNG0,SN06}\,\footnote{A large gluon component of the $\sigma$ meson with a Breit-Wigner mass and width  of about 1 GeV has been also advocated in \cite{OCHS}. A large gluon component for the $\sigma$ and $f_0(980)$ also emerges from a calculation of the gluonia spectrum using a Nambu-Jona-Lasinio model \cite{FRASCA}.}. The averaged result  of the $\gamma\gamma$ direct width of about 1.2(3) keV (see Eq.   \ref{eq:resdir2}), when  runned at the on-shell $\sigma$ mass, can indicate an eventual  large gluon component of the $\sigma$ \footnote{An isoscalar $S_2\equiv \ga{1/ \sqrt{2}}\dr\ga\bar uu+\bar dd\dr$ state is expected to have a $\gamma\gamma$ width of about 4 keV \cite{SNG0,SN06} from QCD spectral sum rules (QSSR) \cite{SVZ,SNB} and quark model \cite{ROSNER}, while a pure gluonium state $\sigma_B$ with $M_{\sigma_B}=1$ GeV has a width of about  (0.2-0.6) keV \cite{VENEZIA,SNG0} from some low-energy theorems.}. Moreover, our result seems to rule out a large four-quark component which has a too small direct $\gamma\gamma$ width of about some few eV \cite{SNA0} from QSSR or from some $\sigma$-like models \cite{ACHASOV}.  \\
{\it -- The $f_0(980)$}: the direct $\gamma\gamma$ width of 0.27 keV for the $f_0(980)$  in Eq.   (\ref{eq:resdir}) is neither compatible with a large four-quark component (too small) nor with a large $S_2$ component (too big). A pure $\bar ss$ component where a $\gamma\gamma$ width is expected to be about 0.4 keV \cite{SNG0,SN06,ANISOVICH} is not however favoured because of the non-zero coupling of $f_0(980)$ to $\pi\pi$ [$|g_{fK^+K^-}|/|g_{f\pi^+\pi^-}|=2.59(1.34)$] obtained in \cite{KMN,MNW}. 
%%%%%%%%%%%%%%%%%%%%%%%%%%%%%%%%
\section{Gluonium production from $J/\psi$ and $\phi$ radiative decays}
%\to\gamma G$ and $\phi\to\gamma G$}
\vspace*{-0.25cm}
\nin
%%%%%%%%%%%%%%%%%%%%%%%%%%%%%%%%%%
%{\it -- Gluonium productions from $J/\psi\to\gamma S$ and $\phi\to\gamma S$} 
The previous possible gluonium assignement of the $\sigma$ and $f_0(980)$ can be tested from the $J/\psi$ and $\phi$ radiative decays. This can be done following the work of \cite{NSVZ2}, using dispersion relation techniques and the Euler-Heisenberg effective Lagrangian, where the gluonic part of the amplitude can be converted into the non-perturbative matrix element $\la 0|\alpha_s G^2|S\ra \sim M_S^2f_S$, where the decay constant $f_S~(S\equiv \sigma_B,~G)$ can be obtained from QSSR which predicts two scalar gluonia ($the \sigma_B (1)$ coupled strongly to $\pi\pi,\bar KK$ due to OZI violations and the G(1.5-1.6) coupled to $\eta'\eta', \eta\eta'$ obtained using quenched lattice)  \cite{VENEZIA,SNG0,SN06}. In this way, one obtains:
\bea
\Gamma(J/\psi\to S\gamma)&\simeq&{\alpha^3\pi\over {\beta_1^2656100}}\ga{M_{J/\psi}\over M_c}\dr^4
\ga{{M_S}\over M_c}\dr^4\times\nnb\\
&&{\ga 1-M^2_S/M^2_{J/\psi}\dr^3\over {\Gamma\ga J/\psi\to e^+e^-\dr}}f^2_S~,
\eea
where $\beta_1=-1/2(11-2n_f/3)$ for $SU(n)_f$ is the first coefficient of the $\beta$ function; $M_c$ is the constituent charm quark mass which we take to be about $M_{J/\psi}/2$. Using the decay constant $f_S\equiv f_{\sigma_B}\simeq ( 1.4\sim 1.0)$ GeV for $M_{\sigma_B}\approx (0.75-1)$ GeV \cite{VENEZIA,SNG0,SN06}, one predicts:
\beq
B(J/\psi\to \sigma_B\gamma)\times B(\sigma_B\to {\rm all})\simeq  (0.4\sim 1.0)\times 10^{-3}~,
\eeq
which one can compare with \cite{PDG}:  $B(J/\psi\to \eta'\gamma)\simeq (4.7\pm 0.27)\times 10^{-3}$ and $B(J/\psi\to f_2\gamma)\simeq (1.43\pm 0.11)\times 10^{-3}$.  Applying this result to the case of the $G(1.5-1.6)$ glueball of higher mass, one obtains \cite{VENEZIA,SNG0,SN06}:
\beq
B(J/\psi\to G\gamma)\simeq  (5.0\pm 3.8)\times 10^{-4}~,
\eeq
where we have used $f_G\simeq (0.39\pm 0.15)$ GeV. This prediction, which needs to be improved,
is in agreement with the experimental lower bound of $ (5.7\pm 0.8)\times 10^{-4}$ \cite{PDG}. 
Extrapolating this analysis to the case of the $\phi$-meson, one obtains by using the strange quark constituent mass $M_s\simeq M_\phi/2$:
\beq
B(\phi\to S\gamma)\simeq 1.2\times 10^{-4}~,
\eeq
which, despite the crude approximation used, compares quite will with the data \cite{PDG}: $B(\phi\to \pi^+\pi^-\gamma)\simeq (0.41\pm 0.1)\times 10^{-4}$ and 
$B(\phi\to f_0(980)\gamma)\simeq (3.22\pm 0.19)\times 10^{-4}$. 
%%%%%%%%%%%%%%%%%%%%%%%%%%%%%%%%
\section{Gluonium production from $D_s$ semi-leptonic decays}
%\to\gamma G$ and $\phi\to\gamma G$}
\vspace*{-0.25cm}
\nin
%%%%%%%%%%%%%%%%%%%%%%%%%%%%%%%%%%
An analogous analysis has been done for $D_{s}$ semileptonic decays, where one expects that the gluonium production will be of similar strength as the one for a $\bar qq$ state \cite{DOSCH}:
\beq
{\Gamma[D_s\to\sigma_B(gg)l\nu]\over \Gamma[D_s\to S_2(\bar qq)l\nu]}\approx {1\over |f_+(0)|^2}\ga f_{\sigma_B}\over M_c\dr^2\simeq {\cal O}(1)~,
\eeq
for $f_{\sigma_B}\approx 1$ GeV, where $|f_+(0)|\simeq 0.5$ \cite{SEMILEP} is the form factor associated to the $\bar qq$ semileptonic production. Some other hadronic processes like $J/\psi\to \omega,\phi +\pi\pi/\bar KK$ (see e.g. \cite{PEAN}) and $D_{(s)},B_{(s)}\to 3\pi,...$ (see e.g. \cite{GATTO}) decays can also be studied, but the mechanism for the decays are expected to be more complex than the one of radiative and semi-leptonic processes discussed previously. However, these analysis emphasize the important r\^ole of the kaon loops contributions, which in our approach are due to the rescattering contributions. We plan to come back to these issues in a future work.
%%%%%%%%%%%%%%%%%%%%%%%%%%%%%%%%
\section{Conclusions}
\vspace*{-0.25cm}
\nin
%%%%%%%%%%%%%%%%%%%%%%%%%%%%%%%%%%
At first sight, a $\bar qq$-gluonium mixing scheme like the one proposed below 1 GeV  in \cite{BN,SNG0} might be appropriate for describing the $\sigma$ and $f_0(980)$ obtained from our fits of $\pi\pi\to\pi\pi/\bar KK$ and $\gamma\gamma\to \pi\pi$ scatterings. The values of their ``direct widths" favour a large gluon content for the $\sigma$ meson but are not decisive for explaining the substructure of the $f_0(980)$ meson. However, the large values of the rescattering widths, due to meson loops because of the large couplings of the $\sigma$ and $f_0(980)$ to $\pi\pi$ or/and $\bar KK$, can be also obtained if they are gluonia states but not necessarily if they are four-quark (diquark-antidiquark) or molecule states as currently claimed  in the existing literature.  The agreement of our predictions for a gluonium production through radiative $\phi$ radiative decays with the data seems to support some large gluon component for the $\sigma$ and to a lesser extent for the $f_0(980)$. This test can be pursued in the analysis of $J/\psi$ radiative and $D_s$ semi-leptonic decays. We plan to analyze in details the substructure of these light scalar mesons by including mixings in a future work.
 %%%%%%%%%%%%%%%%%%%%%%%%%%%%%%%%
\section*{Acknowledgements}
\vspace*{-0.25cm}
\nin
%%%%%%%%%%%%%%%%%%%%%%%%%%%%%%%%%%
We thank Eduardo de Rafael and Jose Oller for communications and discussions. We also thank Wolfgang Ochs for comments on the preliminary draft and for several email exchanges. This work has been partly supported by CNRS-IN2P3 within the project Non-perturbative QCD and Hadron Physics. X.G. Wang thanks the Laboratoire de Physique Th\'eorique et Astroparticules (LPTA) of Montpellier for the hospitality. 
%%%%%%%%%%%%%%%%%%
%\vfill\eject
%%%%%%%%%%%%%%%

%%%%%%%%%%%%%%%%%%%
\end{document}